\documentclass[aps,prb,twocolumn,showpacs]{revtex4}
\usepackage{graphicx} %Include figure files
\usepackage{amsmath}
\usepackage{color}
\newcommand{\be}{\begin{equation}}
\newcommand{\ee}{\end{equation}}
\newcommand{\bea}{\begin{eqnarray}}
\newcommand{\eea}{\end{eqnarray}}
\newcommand{\bse}{\begin{subequations}}
\newcommand{\ese}{\end{subequations}}

\begin{document}

\title{Elaboration of the $\alpha$-Model Derived from the BCS Theory of Superconductivity}
\author{D. C. Johnston}
\altaffiliation{johnston@ameslab.gov} 
\affiliation {Ames Laboratory and Department of Physics and Astronomy, Iowa State University, Ames, Iowa 50011, USA}

\date{\today}

\begin{abstract}

The single-band $\alpha$-model of superconductivity [H. Padamsee, et al., J. Low Temp. Phys. {\bf 12}, 387 (1973)] is a popular model that was adapted from the single-band Bardeen-Cooper-Schrieffer (BCS) theory of superconductivity mainly to allow fits to electronic heat capacity versus temperature $T$ data that deviate from the BCS prediction.  The model assumes that the normalized superconducting order parameter $\Delta(T)/\Delta(0)$ and therefore the normalized London penetration depth $\lambda_{\rm L}(T)/\lambda_{\rm L}(0)$ are the same as in BCS theory, calculated using the BCS value $\alpha_{\rm BCS}\approx 1.764$ of $\alpha\equiv \Delta(0)/k_{\rm B}T_{\rm c}$, where $k_{\rm B}$ is Boltzmann's constant and $T_{\rm c}$ is the superconducting transition temperature.  On the other hand, to calculate the electronic free energy, entropy, heat capacity and thermodynamic critical field versus $T$, the $\alpha$-model takes $\alpha$ to be an adjustable parameter.  Here we write the BCS equations and limiting behaviors for the superconducting state thermodynamic properties explicitly in terms of~$\alpha$, as needed for calculations within the $\alpha$-model, and present  plots of the results  versus $T$ and $\alpha$ that are compared with the respective BCS predictions.  Mechanisms such as gap anisotropy and strong coupling that can cause deviations of the thermodynamics from the BCS predictions, especially the heat capacity jump at $T_{\rm c}$, are considered.  Extensions of the $\alpha$-model that have appeared in the literature such as the two-band model are also discussed.  Tables of values of $\Delta(T)/\Delta(0)$, the normalized London parameter $\Lambda(T)/\Lambda(0)$ and $\lambda_{\rm L}(T)/\lambda_{\rm L}(0)$ calculated from the BCS theory using $\alpha = \alpha_{\rm BCS}$ are provided, which are the same in the $\alpha$-model by assumption.  Tables of values of the entropy, heat capacity and thermodynamic critical field versus $T$ for seven values of $\alpha$ including $\alpha_{\rm BCS}$ are also presented.

\end{abstract}


\maketitle

\section{Introduction}

The 1957 Bardeen-Cooper-Schrieffer (BCS) microscopic single-band clean-limit mean-field theory of superconductivity describes superconductivity as arising from an indirect attractive interaction between Cooper pairs of electrons mediated by the electron-phonon interaction.\cite{Bardeen1957}  In the BCS theory, a Cooper pair has spin~$S=0$ (a spin singlet) with zero angular momentum corresponding to $s$-wave superconductivity.  The BCS theory is a weak-coupling theory, which means that
\be
\frac{T_{\rm c}}{\Theta_{\rm D}}\ll1\quad{\rm and}\quad \frac{\Delta(0)}{k_{\rm B}\Theta_{\rm D}}\ll1,
\label{Eq:WeakCoupDef}
\ee
where $T_{\rm c}$ is the superconducting transition temperature, $\Theta_{\rm D}$ is the Debye temperature, $k_{\rm B}\Theta_{\rm D}$ is the maximum phonon energy within the Debye theory,\cite{Kittel2005} $k_{\rm B}$ is Boltzmann's constant, and $\Delta(T)$ is the superconducting order parameter versus temperature~$T$, which is the activation energy for single quasiparticle (electron and hole) excitations of the superconducting ground state and is often just called the superconducting gap.  The actual superconducting energy gap $2\Delta$ is centered on the Fermi energy of the metal and is thus qualitatively different from the energy gap in a semiconductor where the energy gap is between the top of the valence band and bottom of the conduction band.  The BCS theory makes precise predictions of $\Delta(T)$ and the magnetic penetration depths derived from it, together with predictions of thermodynamic quantities which include the superconducting state electronic free energy $F_{\rm es}(T)$, thermodynamic critical field $H_{\rm c}(T)$, entropy $S_{\rm es}(T)$ and heat capacity~$C_{\rm es}(T)$.    

The BCS theory predicts that $\Delta(T)$ decreases continuously and monotonically to zero as $T_{\rm c}$ is approached from below, resulting in a second-order phase transition at $T_{\rm c}$.  A finite discontinuous increase (``jump'') $\Delta C_{\rm e}(T_{\rm c}) = C_{\rm es}(T_{\rm c}) - C_{\rm en}(T_{\rm c})$ in $C_{\rm e}(T)$ occurs on entering the superconducting state from the normal state with decreasing $T$\@.  The jump is given in the weak-coupling limit by $\Delta C_{\rm e}/\gamma_{\rm n} T_{\rm c}\approx 1.43$ for all BCS superconductors because it is a law of corresponding states, where the normal-state electronic heat capacity is $C_{\rm en}(T) = \gamma_{\rm n}T$ and $\gamma_{\rm n}$ is called the Sommerfeld electronic heat capacity coefficient.  This value of $\Delta C_{\rm e}/\gamma_{\rm n} T_{\rm c}$ was confirmed for some superconductors such as Al and Ga, but other superconductors showed $\Delta C_{\rm e}/\gamma_{\rm n} T_{\rm c} > 1.43$ such as the value 2.7 for Pb.\cite{Meservey1969}  The enhanced jumps were subsequently determined to arise in superconductors that are not in the weak-coupling limit, termed moderate- or strong-coupling superconductors.\cite{Carbotte1990}  There exists no rigorous theoretical expression for $C_{\rm es}(T)$ for such superconductors that can be routinely used to fit experimental heat capacity data such as the heat capacity jump at $T_{\rm c}$.

To provide a model for fitting experimental $C_{\rm es}(T)$ data such as $\Delta C_{\rm e}(T_{\rm c})/\gamma_{\rm n} T_{\rm c}$ and other thermodynamic properties for moderate- and strong-coupling superconductors, Padamsee, Neighbor and Shiffman introduced the single-band \mbox{$\alpha$-model} in 1973,\cite{Padamsee1973} which is able to fit $C_{\rm es}(T)$ data not only for strong-coupling superconductors with $\Delta C_{\rm e}(T_{\rm c})/\gamma_{\rm n} T_{\rm c} > 1.43$ which was the original motivation, but also for superconductors with $\Delta C_{\rm e}(T_{\rm c})/\gamma_{\rm n} T_{\rm c} < 1.43$ which can arise from anisotropy of $\Delta$ in momentum space (see, e.g., Ref.~\onlinecite{Openov2004}, and Sec.~\ref{Sec:AnisOrdPars} below).  The $\alpha$-model assumes that the normalized gap $\Delta(T)/\Delta(0)$ and therefore the normalized London penetration depth $\lambda_{\rm L}(T)/\lambda_{\rm L}(0)$ are the same as in the BCS theory which are calculated using the BCS value $\alpha_{\rm BCS} \approx 1.764$ of $\alpha \equiv \Delta(0)/k_{\rm B}T_{\rm c}$.  On the other hand, the normalized superconducting state $S_{\rm es}(T)/\gamma_{\rm n} T_{\rm c}$, $C_{\rm es}(T)/\gamma T_{\rm c}$, $F_{\rm es}(T)/\gamma T_{\rm c}^2$ and $H_{\rm c}(T)/\sqrt{\gamma T_{\rm c}^2}$ are calculated taking~$\alpha$ to be an adjustable parameter in the corresponding BCS equations to allow fits to experimental data that deviate from the BCS predictions.

Thus the $\alpha$-model is not self-consistent, but provides a popular model with which experimentalists can fit their electronic superconducting state thermodynamic data that deviate from the BCS predictions and to quantify those deviations.  However, few of the many papers reporting use of the $\alpha$-model to fit experimental $C_{\rm es}(T)$ data explain how the theoretical values for the fits were obtained.  Indeed, the original exposition by Pademsee et~al.\ in Ref.~\onlinecite{Padamsee1973} is not completely developed.  The main purpose of the present paper is to write the BCS equations for the thermodynamic properties explicitly in terms of~$\alpha$ so that these equations can be directly used to numerically calculate and plot these properties versus both $T$ and $\alpha$.  We also discuss predictions of the BCS theory itself for comparison with the predictions of the $\alpha$-model.  Tables of values of the various superconducting state properties for both the BCS theory versus $T$ and the $\alpha$-model versus $T$ and $\alpha$ are provided.

In order to introduce and explain calculations needed for the $\alpha$-model, the predictions of the BCS theory\cite{Bardeen1957,Tinkham1975} and their limiting behaviors at low and high temperatures are first discussed.  These BCS superconducting state predictions are described in Secs.~\ref{Sec:BCSGap}--\ref{BCSPenDepth} including plots of calculated data. The equations are written in terms of the $\alpha$~parameter of the $\alpha$-model, which are then used in Sec.~\ref{Sec:AlphaModel} to calculate $S_{\rm es}$, $C_{\rm es}$, and $H_{\rm c}$ versus $T$ and~$\alpha$, which are compared with the BCS theory predictions.  The role of gap anisotropy as a mechanism for producing reduced heat capacity jumps $\Delta C_{\rm e}/\gamma_{\rm n} T_{\rm c} < 1.43$ is discussed and calculated for several example cases in Sec.~\ref{Sec:AnisOrdPars}.   Extensions of the single-band $\alpha$-model that have appeared in the literature are briefly discussed in Sec.~\ref{Sec:Extensions}.  A summary is given in Sec.~\ref{Summary}.  In this paper, we use Gaussian cgs units throughout.\cite{Bardeen1957}  

In the Appendix, tables are provided of values of $\Delta(t)/\Delta(0)$, the London parameter $\Lambda(t)/\Lambda(0)$ and $\lambda_{\rm L}(t)/\lambda_{\rm L}(0)$  calculated from the BCS theory using $\alpha = \alpha_{\rm BCS}$, where $t=T/T_{\rm c}$, which are the same in the $\alpha$-model by assumption, and a table of the BCS prediction of the Pippard penetration depth $\lambda_{\rm P}(t)/\lambda_{\rm P}(0)$ versus $t$.  Also provided in the Appendix are tables of the normalized values $S_{\rm es}(t)/\gamma_{\rm n}T_{\rm c}$, $C_{\rm es}(t)/\gamma_{\rm n}T_{\rm c}$  and $H_{\rm c}(t)/\sqrt{\gamma_{\rm n}T_{\rm c}^2}$ versus~$t$ for seven values of $\alpha$ including $\alpha_{\rm BCS}$.  These tables supplement the very popular table in Ref.~\onlinecite{Muhlschlegel1959} of superconducting state properties versus temperature predicted by the BCS theory.  In the areas of overlap, the values in our tables agree with those in Ref.~\onlinecite{Muhlschlegel1959}.\cite{MuhlError}

\section{\label{Sec:BCSGap} BCS Gap Equation and $T_{\rm c}$}

The BCS gap equation is
\bse
\label{Eqs:BCSGap}
\be
\int_0^{k_{\rm B}\Theta_{\rm D}} \frac{d\epsilon}{E}\tanh\left(\frac{E}{2k_{\rm B}T}\right) = \frac{1}{N(0)V},
\label{Eq:BCSGap}
\ee
where
\be
E = \sqrt{\epsilon^2 + \Delta^2}
\label{Eq:QPEnergy}
\ee
\ese
is the energy of an electron excited above the superconducting energy gap and of a simultaneously excited hole below the gap with respect to the Fermi energy $E_{\rm F}$ at the center of the gap, both of which are defined to be positive, $\epsilon$ is the normal-state single-particle energy, $V$ is the electron-phonon coupling constant and $\Delta$ is assumed to be the same everywhere on the Fermi surface (isotropic $s$-wave superconductivity).  The excited electrons and holes are together termed quasiparticles.  It is assumed that $k_{\rm B}\Theta_{\rm D}/E_{\rm F}\ll 1$, where $E_{\rm F}$ is the Fermi energy measured from the bottom of the electron conduction band. In the BCS theory, the energies $\epsilon$ and $E$ are measured with respect to $E_{\rm F}$, so $E_{\rm F} \equiv 0$ for those energies.  $N(0) \equiv N(E_{\rm F})$ is the normal-state electronic density of states at $\epsilon = E_{\rm F} = 0$ for a single spin direction, which is assumed to be constant over the energy range of interest in the theory.  The full energy gap that develops below the superconducting transition temperature $T_{\rm c}$ is $2\Delta(T)$, with $E_{\rm F}$ in the middle of the gap.  Thus $\Delta$ is the activation energy for a quasiparticle excitation and $2\Delta$ is the energy required to break up a Cooper pair via simultaneous excitations of an electron and a hole quasiparticle.  The energy $E$ in Eq.~(\ref{Eq:QPEnergy}) approaches the normal state energy $\epsilon$ for $\epsilon \gg \Delta$.

To determine $T_{\rm c}$, one sets $\Delta =0$ and $T=T_{\rm c}$ in Eqs.~(\ref{Eqs:BCSGap}) and takes the weak-coupling limit $T_{\rm c}/\Theta_{\rm D}\ll 1$, yielding the BCS equation for $T_{\rm c}$ given by
\bea
T_{\rm c} &=& \left(\frac{2e^{\gamma_{\rm E}}}{\pi}\right)\Theta_{\rm D}\exp\left(-\frac{1}{N(0)V}\right)\label{Eq:TcBCS}\\*
&\approx&1.134\ \Theta_{\rm D}\exp\left(-\frac{1}{N(0)V}\right),\nonumber
\eea
where $\gamma_{\rm E}\approx 0.5772$ is Euler's constant.  Defining the symbol
\be
\alpha_{\rm BCS} \equiv \frac{\pi}{ e^{\gamma_{\rm E}} }\approx 1.764,
\label{Eq:alphaBCS}\\*
\ee
Eq.~(\ref{Eq:TcBCS}) can be written
\be
\frac{1}{N(0)V} = \ln\left(\frac{2\Theta_{\rm D}}{\alpha_{\rm BCS}T_{\rm c}}\right).
\ee
Inserting this into the right side of Eq.~(\ref{Eq:BCSGap}) gives a second form of the gap equation
\be
\int_0^{k_{\rm B}\Theta_{\rm D}} \frac{d\epsilon}{E}\tanh\left(\frac{E}{2k_{\rm B}T}\right) = \ln\left(\frac{2\Theta_{\rm D}}{\alpha_{\rm BCS}T_{\rm c}}\right).
\label{Eq:BCSGap55}
\ee

\section{BCS Superconducting Order Parameter $\Delta(T)$}

Taking the limit $T\to0$ and the weak-coupling limit $T_{\rm c}/\Theta_{\rm D}\to 0$ in Eq.~(\ref{Eq:BCSGap55}) allows $\Delta(0)$ to be evaluated as
\bse
\be
\frac{\Delta(0)}{k_{\rm B}T_{\rm c}} = \alpha_{\rm BCS}.
\label{Eq:alphBCSDelt0}
\ee
The full superconducting energy gap at $T=0$ is therefore
\be
\frac{2\Delta(0)}{k_{\rm B}T_{\rm c}} = 2\,\alpha_{\rm BCS}= \frac{2\pi}{ e^{\gamma_{\rm E}} }\approx 3.528.
\label{Eq:BCSGapValue}\\*
\ee
\ese
Inserting the expression for $\alpha_{\rm BCS}$ in Eq.~(\ref{Eq:alphBCSDelt0}) into the right side of the gap equation~(\ref{Eq:BCSGap55}) gives a third form of the gap equation
\be
\int_0^{k_{\rm B}\Theta_{\rm D}} \frac{d\epsilon}{E}\tanh\left(\frac{E}{2k_{\rm B}T}\right) = \ln\left[\frac{2k_{\rm B}\Theta_{\rm D}}{\Delta(0)}\right].
\label{Eq:BCSGap66}
\ee

In order to calculate the temperature dependence of $\Delta$ we define dimensionless reduced variables
\be
\tilde{\Delta} = \frac{\Delta}{\Delta(0)},\qquad \tilde{\epsilon} = \frac{\epsilon}{\Delta(0)},\qquad t = \frac{T}{T_{\rm c}}.
\label{Eq:RedVars}
\ee
From Eq.~(\ref{Eq:QPEnergy}), the excited quasiparticles have reduced energy
\be
\tilde{E} = \frac{E}{\Delta(0)} = \sqrt{\tilde{\epsilon}^2+\tilde{\Delta}^2}.
\label{Eq:EbarDef}
\ee
Using Eq.~(\ref{Eq:alphBCSDelt0}) and the definitions~(\ref{Eq:RedVars}) and~(\ref{Eq:EbarDef}), the gap equation~(\ref{Eq:BCSGap66}) expressed in dimensionless variables is
\be
\int_0^{\frac{k_{\rm B}\Theta_{\rm D}}{\Delta(0)}} \frac{d\tilde{\epsilon}}{\tilde{E}}\tanh\left(\frac{\alpha_{\rm BCS}\tilde{E}}{2t}\right) = \ln\left[\frac{2k_{\rm B}\Theta_{\rm D}}{\Delta(0)}\right],
\label{Eq:BCSGap5}
\ee
where now the ratio $k_{\rm B}\Theta_{\rm D}/\Delta(0)$ occurs both in the upper limit to the integral on the left side and in the argument of the logarithm on the right side.

\begin{figure}
\includegraphics[width=3.in]{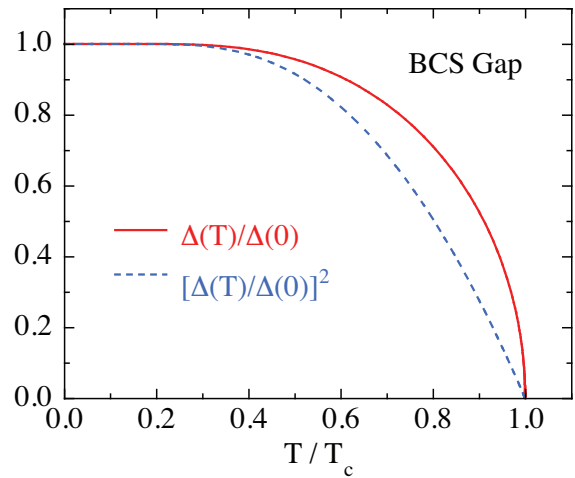}
\caption{(Color online) Normalized BCS energy gap $\Delta(T)/\Delta(0)$ and its square versus reduced temperature $T/T_{\rm c}$ as obtained by numerically solving Eq.~(\ref{Eq:BCSGap5}) in the weak-coupling limit~(\ref{Eq:WeakCoupDef}).}
\label{Fig:BCSGapVsT_AB} 
\end{figure}

Numerically solving Eqs.~(\ref{Eq:EbarDef}) and~(\ref{Eq:BCSGap5}) for $\tilde{\Delta}$ at fixed values of $t$ in the asymptotic weak-coupling regime with $k_{\rm B}\Theta_{\rm D}/\Delta(0) = 500\gg1$, where the calculated $\tilde{\Delta}(t)$ is numerically independent of the precise value of $k_{\rm B}\Theta_{\rm D}/\Delta(0)$, yields the $\tilde{\Delta}$ versus~$t$ data shown in Fig.~\ref{Fig:BCSGapVsT_AB}.  A list of $\tilde{\Delta}(t)$ values is given in Table~\ref{Tab:BCSGap} in the Appendix, where a logarithmic scale in $1-t$ is used to provide a high density of $\tilde{\Delta}$ values for $t\to1$ where $\tilde{\Delta}$ varies rapidly.  Also shown in Fig.~\ref{Fig:BCSGapVsT_AB} is a plot of $\tilde{\Delta}^2(t)$ versus $t$, from which one sees that the order parameter near $T_{\rm c}$ is $\Delta(t\to 1)\sim \sqrt{1-t}$, a $t$ dependence characteristic of mean-field behavior as further discussed below.

The low-$t$ behavior of $\tilde{\Delta}(t)$ can be obtained from Eqs.~(\ref{Eq:EbarDef}) and~(\ref{Eq:BCSGap5}) by doing an integration by parts and then utilizing the weak-coupling limit $k_{\rm B}\Theta_{\rm D}/\Delta(0) \to \infty$, yielding
\bea
\tilde{\Delta}(t\to0) &=& 1-\sqrt{\frac{2\pi t}{\alpha_{\rm BCS}}}\bigg[1-2\frac{t}{\alpha_{\rm BCS}} + 8\left(\frac{t}{\alpha_{\rm BCS}}\right)^2\nonumber\\*
&&-\ 48\left(\frac{t}{\alpha_{\rm BCS}}\right)^3 + 384\left(\frac{t}{\alpha_{\rm BCS}}\right)^4 +{\cal O}(t^5) \bigg]\nonumber\\*
&& \times\ e^{-\alpha_{\rm BCS}/t},
\label{Eq:DeltaLowt}
\eea
from which expressions for the temperature derivatives $d^n\tilde{\Delta}(t\to0)/dt^n$ can also be calculated.

An expression for computing $d\tilde{\Delta}^2/dt$ for $0\leq t\leq 1$, which we need below to calculate $C_{\rm es}(t)$, is obtained by taking the $t$ derivative of the gap equation~(\ref{Eq:BCSGap5}) and solving for $d\tilde{\Delta}^2/dt$, yielding
\bse
\label{Eqs:dDelta2/dt}
\be
\frac{d\tilde{\Delta}^2(t)}{dt} = \frac{\int_0^\infty {\rm sech}^2(g)d\tilde{\epsilon}}{\int_0^\infty\left[\frac{t\,{\rm sech}^2(g)}{2\left(\tilde{\epsilon}^2+\tilde{\Delta}^2\right)}-\frac{t^2\tanh(g)}{\alpha_{\rm BCS}\left(\tilde{\epsilon}^2+\tilde{\Delta}^2\right)^{3/2}}\right]d\tilde{\epsilon}},
\label{Eq:dD2dt}
\ee
where
\be
g\equiv \frac{\alpha_{\rm BCS}\sqrt{\tilde{\epsilon}^2+{\tilde{\Delta}^2}}}{2t}.
\ee
\ese
Alternatively, one can use Eqs.~(\ref{Eqs:I(t)}) and~(\ref{Eq:GapFromLambda}) below to calculate $d\tilde{\Delta}^2(t)/dt$, which give the same numerical results as Eqs.~(\ref{Eqs:dDelta2/dt}).

For $T\to T_{\rm c}$, one obtains
\bea
\frac{d\tilde{\Delta}^2(t)}{dt}\bigg|_{t\to1} &=& -\frac{8e^{2\gamma_{\rm E}}}{7\zeta(3)} \approx -3.016
\label{Eq:dD2dtTc}\\*
&\equiv& -A\nonumber,
\eea
where $\zeta(x)$ is the Riemann zeta function.  From Fig.~\ref{Fig:BCSGapVsT_AB} and Eq.~(\ref{Eq:dD2dtTc}), for $t\to1$ one has
\be
\tilde{\Delta}^2(t\to1) = A(1-t).
\label{Eq:Delta2tto1}
\ee 
Equations~(\ref{Eq:dD2dtTc}) and~(\ref{Eq:Delta2tto1}) give
\bse
\bea
\tilde{\Delta}(t\to1) &=& \sqrt{A}\ \sqrt{1-t}\label{Eq:Deltatto1}\\*
&\approx& 1.737\,\sqrt{1-t},\nonumber
\eea
which is a $t$ dependence of the order parameter characteristic of mean-field theories as noted above. Equation~(\ref{Eq:Deltatto1}) can also be written as
\bea
\frac{\Delta(t\to 1)}{k_{\rm B}T_{\rm c}} &=& \alpha_{\rm BCS}\tilde{\Delta}(t\to1) = \sqrt{\frac{8\pi^2}{7\zeta(3)}}\ \sqrt{1-t}\nonumber\\*
 &\approx& 3.063\,\sqrt{1-t},
\eea
\ese
where we used the expression for $\alpha_{\rm BCS}$ in Eq.~(\ref{Eq:alphaBCS}).  For later use, we combine Eqs.~(\ref{Eq:dD2dtTc}) and~(\ref{Eq:Delta2tto1}) to obtain
\be
\frac{t}{\tilde{\Delta}^2}\,\frac{d\tilde{\Delta}^2}{dt}\Big|_{t\to1} = -\frac{1}{1-t}
\label{Eq:limtto1D}
\ee

Using Eqs.~(\ref{Eqs:dDelta2/dt}), $d\tilde{\Delta}^2/dt$ and $d\tilde{\Delta}/dt$ at $t\to0$ are calculated to be
\bse
\bea
\frac{d\tilde{\Delta}^2(t\to0)}{dt} &=& -\frac{2\sqrt{2\pi\alpha_{\rm BCS}}}{t^{3/2}}\,e^{-\alpha_{\rm BCS}/t}\label{Eq:dD2dtLowT}\\*
\frac{d\tilde{\Delta}(t\to0)}{dt} &=& \frac{1}{2\tilde{\Delta}}\frac{d\tilde{\Delta}^2(t\to0)}{dt} = \frac{1}{2}\frac{d\tilde{\Delta}^2(t\to0)}{dt}\nonumber\\*
 &=& -\frac{\sqrt{2\pi\alpha_{\rm BCS}}}{t^{3/2}}\,e^{-\alpha_{\rm BCS}/t}.\label{Eq:dDdtLowT}
\eea
\ese
These two results are the same as obtained directly from Eq.~(\ref{Eq:DeltaLowt}) by taking the respective derivative and utilizing only the leading term $\sim \sqrt{t}$ in the prefactor of the exponential.

\section{\label{Sec:SesCes} BCS Electronic Entropy and Heat Capacity}

The normal-state Sommerfeld electronic specific heat coefficient $\gamma_{\rm n}$ is related to $N(0)$ by\cite{Kittel2005}
\be
\gamma_{\rm n} = \frac{2\pi^2k_{\rm B}^2}{3}\,N(0).
\label{Eq:gammaN(0)}
\ee
The $\gamma_{\rm n}$ is the measured value and therefore both $\gamma_{\rm n}$ and $N(0)$ contain the same  enhancements by many-body electron-electron correlations and the electron-phonon interaction.  Using Eq.~(\ref{Eq:gammaN(0)}), the superconducting-state electronic entropy $S_{\rm es}$ and heat capacity $C_{\rm es}$ are expressed in terms of $\gamma_{\rm n}T_{\rm c}$ and reduced variables according to
\bse
\label{Eqs:SesCes}
\be
\frac{S_{\rm es}(t)}{\gamma_{\rm n}T_{\rm c}} = \frac{6\alpha_{\rm BCS}^2}{\pi^2t}\int_0^\infty f(\alpha_{\rm BCS},\tilde{E},t)\left(\tilde{E} + \frac{\tilde{\epsilon}^2}{\tilde{E}}\right)d\tilde{\epsilon},
\label{Eq:Se6}
\ee
\be
\frac{C_{\rm es}(t)}{\gamma_{\rm n}T_{\rm c}} = \frac{6\alpha_{\rm BCS}^3}{\pi^2t}\int_0^\infty f(1-f)\left(\frac{\tilde{E}^2}{t}-\frac{1}{2}\,\frac{d\tilde{\Delta}^2}{dt}\right)d\tilde{\epsilon},
\label{Eq:Ces2}
\ee
where the Fermi-Dirac distribution function is (with $E_{\rm F}=0$)
\bea
f\equiv f(\alpha_{\rm BCS},\tilde{E},t) &=& \frac{1}{e^{\alpha_{\rm BCS}\tilde{E}/t} + 1}\label{Eq:FermiFcnRed}\\*
{\rm with}\ \ \ \frac{E}{k_{\rm B}T} &=&  \frac{\alpha_{\rm BCS}\tilde{E}}{t} .\nonumber
\eea
\ese

\begin{figure}
\includegraphics[width=3.4in]{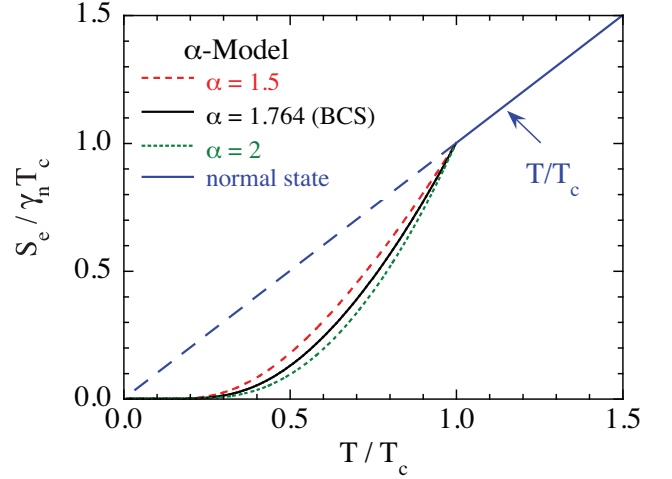}
\caption{(Color online) Normalized electronic entropy $S_{\rm e}(T)/\gamma_{\rm n}T_{\rm c}$ versus reduced temperature $t = T/T_{\rm c}$ in both the superconducting ($T/T_{\rm c}\leq1$) and normal ($T/T_{\rm c}\geq1$) states for three values of $\alpha\equiv \Delta(0)/k_{\rm B}T_{\rm c}$ including the BCS value $\alpha = \alpha_{\rm BCS}$ in Eq.~(\ref{Eq:alphaBCS}). The superconducting state data are calculated using Eq.~(\ref{Eq:Se6}).  The normal-state behavior at $T\geq T_{\rm c}$ is given by $S_{\rm en}/\gamma_{\rm n}T_{\rm c} = T/T_{\rm c}$ and the extrapolation to lower $T$ is the long-dashed line.}
\label{Fig:Alpha_Model_Ses2} 
\end{figure}

\begin{figure}
\includegraphics[width=3.3in]{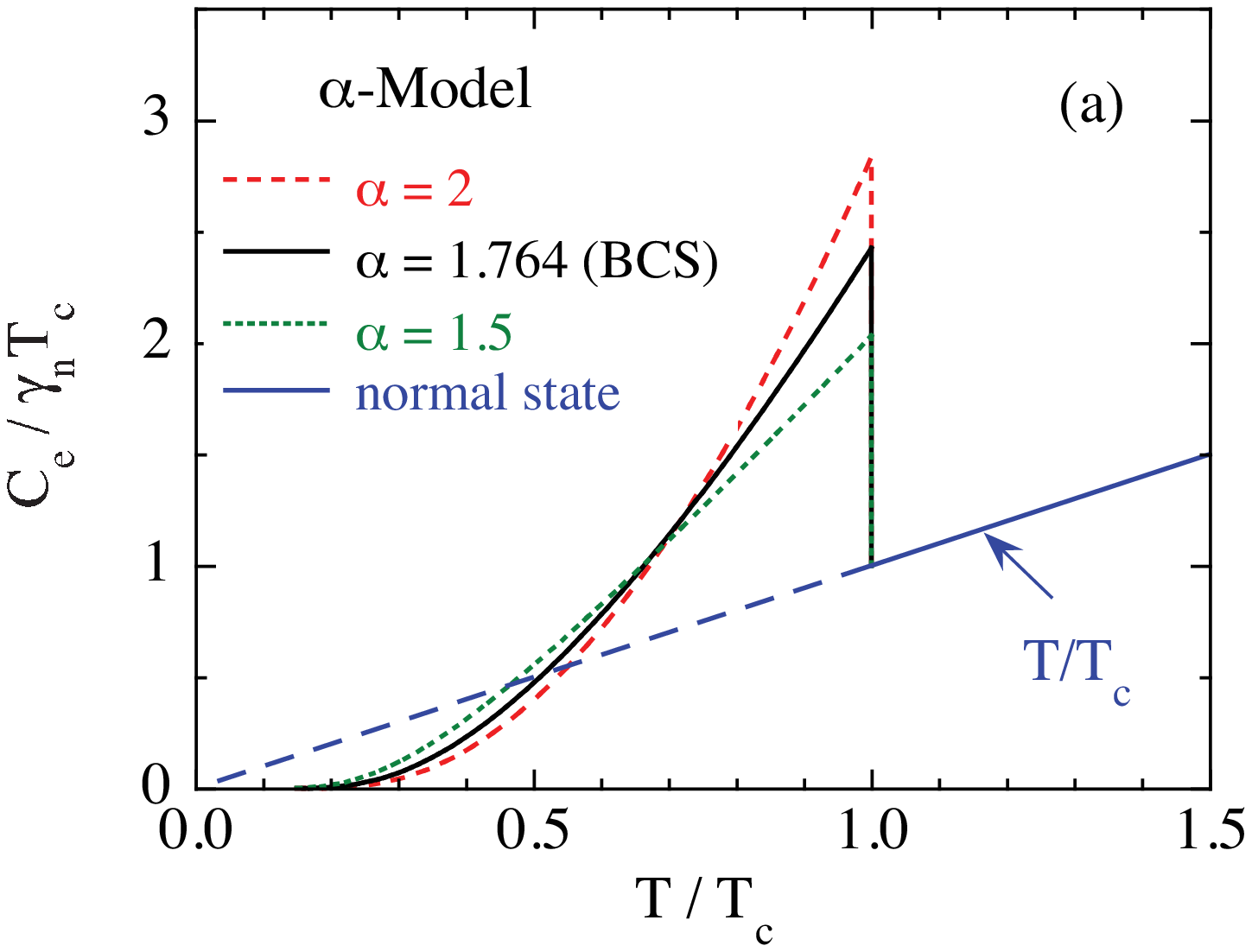}
\includegraphics[width=3.3in]{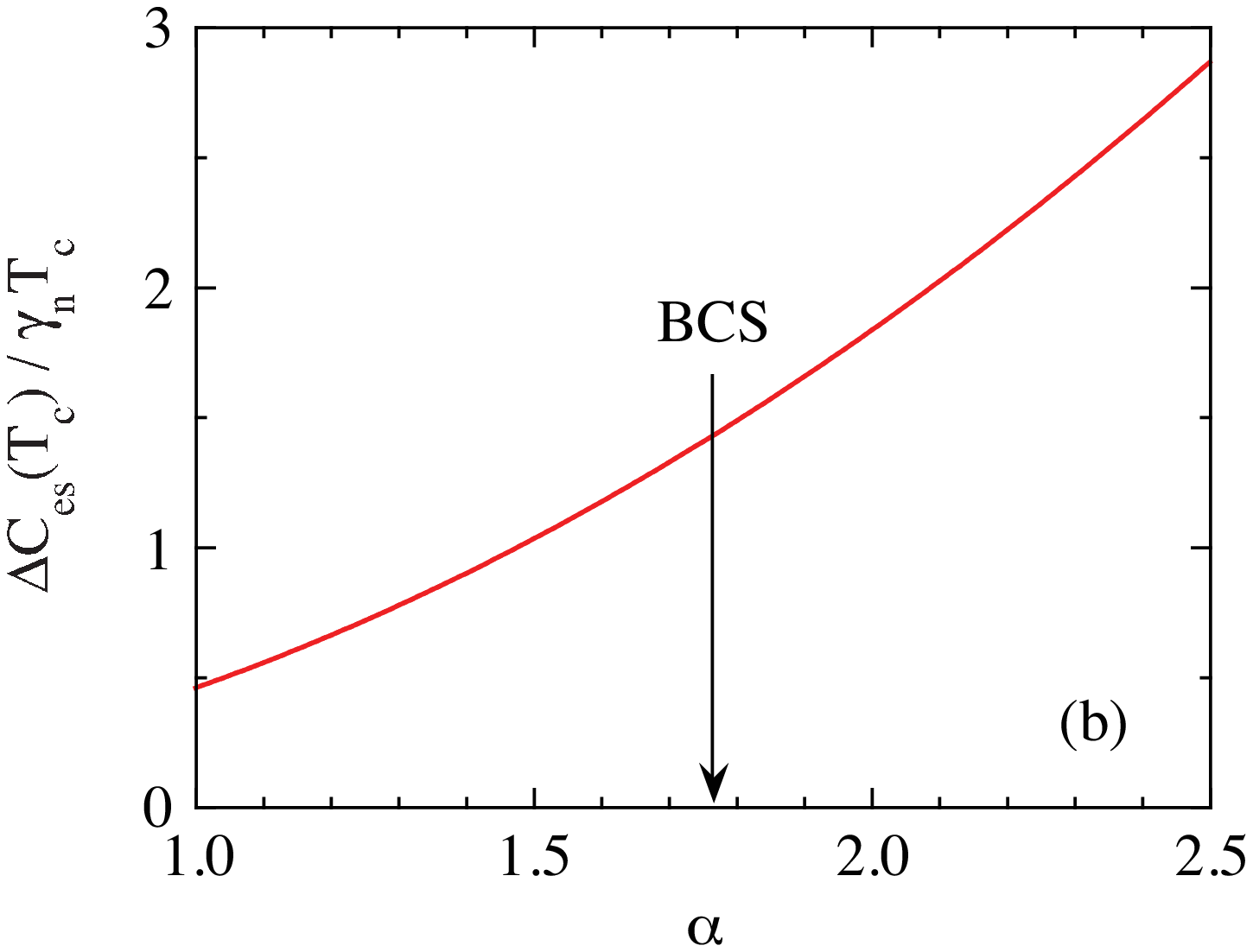}
\caption{(Color online) Calculations using the $\alpha$-model.  (a)~Normalized electronic heat capacity $C_{\rm e}(T)/\gamma_{\rm n}T_{\rm c}$ versus temperature $T$ in both the superconducting ($T/T_{\rm c}\leq1$) and normal ($T/T_{\rm c}\geq1$) states for three values of $\alpha\equiv \Delta(0)/k_{\rm B}T_{\rm c}$ including the BCS value $\alpha = \alpha_{\rm BCS}$ in Eq.~(\ref{Eq:alphaBCS}). The superconducting state data are calculated using Eq.~(\ref{Eq:Ces2}).  The normal-state behavior at $T\geq T_{\rm c}$ is given by $C_{\rm en}/\gamma_{\rm n}T_{\rm c} = T/T_{\rm c}$ and the extrapolation to lower $T$ is the long-dashed line.  (b) Dependence of the specific heat jump $\Delta C_{\rm e}(T_{\rm c})/\gamma_{\rm n}T_{\rm c}$ at $T_{\rm c}$ on~$\alpha$.  The BCS value $\alpha_{\rm BCS}\approx 1.764$ in Eq.~(\ref{Eq:alphaBCS}) is indicated.}
\label{Fig:Alpha_Model_Ces} 
\end{figure}\

The normalized electronic entropy $S_{\rm e}/\gamma_{\rm n}T_{\rm c}$ versus reduced temperature $t$ calculated numerically using Eq.~(\ref{Eq:Se6}) is plotted as the solid black curve in Fig.~\ref{Fig:Alpha_Model_Ses2}.  One sees that the superconducting and normal state entropies are the same at $T_{\rm c}$, which demonstrates that  the superconducting transition is second order with no latent heat. The superconducting state entropy is lower than the normal state entropy for $0\leq t < 1$, showing that the superconducting state is more ordered than the normal state in this temperature range.  The superconducting-state entropy approaches zero exponentially for $t\to0$ due to the superconducting energy gap for quasiparticle excitations.

To obtain $C_{\rm es}$ at a particular $t$ one first determines $\tilde{\Delta}$ at that $t$ using Eq.~(\ref{Eq:BCSGap5}), inserts that into Eqs.~(\ref{Eqs:dDelta2/dt}) to determine $d\tilde{\Delta}^2/dt$ at that $t$, and then inserts these two quantities into Eq.~(\ref{Eq:Ces2}) and does the integral there.  All integrals are done numerically.  A plot of the BCS prediction of $C_{\rm e}$ versus $t$ for $\alpha=\alpha_{\rm BCS}$ is shown as the black solid curve in Fig.~\ref{Fig:Alpha_Model_Ces}(a).  The heat capacity jump $\Delta C_{\rm e}(T_{\rm c})$ on cooling below $T_{\rm c}$ is given by Eq.~(\ref{Eq:Ces2}) as
\be
\frac{\Delta C_{\rm e}(T_{\rm c})}{\gamma_{\rm n}T_{\rm c}} = -\left(\frac{3\alpha_{\rm BCS}{^2}}{2\pi^2}\right)\frac{d\tilde{\Delta}^2}{dt}\bigg|_{t=1}.
\label{Eq:DCesGamTcA}
\ee
Substituting the expressions for $\alpha_{\rm BCS}$ from Eq.~(\ref{Eq:alphaBCS}) and $d\tilde{\Delta}^2/dt|_{t=1}$ from Eq.~(\ref{Eq:dD2dtTc}) into~(\ref{Eq:DCesGamTcA}) gives
\be
\frac{\Delta C_{\rm e}(T_{\rm c})}{\gamma_{\rm n}T_{\rm c}} =\frac{12}{7\zeta(3)} \approx 1.426.
\label{Eq:DCesGamTc}
\ee

At low temperatures $t\lesssim 0.2$ where $\tilde{\Delta}\approx 1$ (see Table~\ref{Tab:BCSGap} in the Appendix), the heat capacity is given by BCS as
\be
\frac{C_{\rm es}(t)}{\gamma_{\rm n}T_{\rm c}} = \frac{3\alpha_{\rm BCS}^3}{2\pi^2t^2}[3K_1(\alpha_{\rm BCS}/t) + K_3(\alpha_{\rm BCS}/t)],
\label{Eq:CesLowT}
\ee
where $K_n(x)$ is the modified Bessel function of the second kind. This heat capacity arises from excitations of electron and hole quasiparticles, and at these low temperatures does not include a contribution from a change with temperature of the superconducting condensation energy $\sim\Delta^2$.  We have verified that numerical data generated using Eq.~(\ref{Eq:Ces2}) are in precise agreement with Eq.~(\ref{Eq:CesLowT}) at $t\lesssim 0.2$.  The expansion of Eq.~(\ref{Eq:CesLowT}) at low $t$ is
\bea
\frac{C_{\rm es}(t)}{\gamma_{\rm n}T_{\rm c}} &=& \frac{3\sqrt{2}\,\alpha_{\rm BCS}}{\pi^{3/2}}\left(\frac{\alpha_{\rm BCS}}{t}\right)^{3/2}\label{Eq:BCSLowT}\\*
& \times& \left[1 + \frac{11}{8}\left(\frac{t}{\alpha_{\rm BCS}}\right) + \frac{225}{128}\left(\frac{t}{\alpha_{\rm BCS}}\right)^2 + {\cal O}(t^3)\right]\nonumber\\*
&\times& e^{-\alpha_{\rm BCS}/t}.\nonumber
\eea
Thus at low $t$, $C_{\rm es}$ decreases exponentially to zero with decreasing $t$.  Using $\alpha_{\rm BCS}/t \equiv \Delta(0)/k_{\rm B}T$, this $T$ dependence is seen to arise from excitations of electron and hole quasiparticles above and below the superconducting energy gap, respectively, with activation energy $\Delta(0)$ for both types of quasiparticle. The numerical prefactor is $3\sqrt{2}\,\alpha_{\rm BCS}/\pi^{3/2}\approx 1.344$.\cite{Bouquet2001Error}

BCS fitted their calculations of $C_{\rm es}(t)/\gamma_{\rm n}T_{\rm c}$ for $t\lesssim0.7$ using Eq.~(\ref{Eq:Ces2}) by the expression $C_{\rm es}(t)/\gamma_{\rm n}T_{\rm c} = a e^{-b/t}$ and obtained $a=8.5$ and $b=1.44$.  Because $a$ was a constant independent of $t$ and the fit was not done in the low-$t$ limit, the fitted value $b=1.44$ in the exponent is not equal to the low-$t$ limit value $\alpha_{\rm BCS} = \Delta(0)/k_{\rm B}T_{\rm c}\approx 1.764$.  The fit was done to compare their theoretical prediction with experimental data that were not in the low-$t$ limit. Within BCS theory, one could evidently obtain $\Delta(0)$ from fits of experimental $C_{\rm es}(T)$ data for $T \lesssim 0.7\, T_{\rm c}$ by the expression $C_{\rm es}(T) = a^\prime e^{-b^\prime/k_{\rm B}T}$ using $\Delta(0)\approx 1.22\, b^\prime$, where $1.22 = 1.76/1.44$.

\section{\label{Sec:FeHc} BCS Electronic Free Energy $F_{\rm es}$ and Thermodynamic Critical Field $H_{\rm c}$}

The electronic Helmholtz free energy $F_{\rm e}$ is\cite{Reif1965} 
\be
F_{\rm e} = U_{\rm e} - TS_{\rm e},
\ee
where $U_{\rm e}$ is the electronic internal energy.  The normal-state free energy is 
\be
\frac{F_{\rm en}(t)}{\gamma_{\rm n}T_{\rm c}^2} = -\frac{t^2}{2}.
\label{Eq:Fen}
\ee
The free energy $F_{\rm es}$ of a BCS superconductor in the weak-coupling limit is
\bea
\frac{F_{\rm es}(t)}{\gamma_{\rm n}T_{\rm c}^2} &=&\label{Eq:FesFen}\\*
&&\hspace{-0.5in}-\frac{3\alpha_{\rm BCS}^2}{\pi^2}\left[\frac{\tilde{\Delta}^2}{4} + \int_0^\infty f(\alpha_{\rm BCS},\tilde{E},t)\left(\frac{2\tilde{\epsilon}^2+\tilde{\Delta}^2}{\tilde{E}}\right)d\tilde{\epsilon}\right].\nonumber
\eea
For $\tilde{\Delta}\to0$ at $t\to1$ one regains the normal-state free energy, i.e.,
\be
\frac{F_{\rm es}(t=1)}{\gamma_{\rm n}T_{\rm c}^2} = \frac{F_{\rm en}(t=1)}{\gamma_{\rm n}T_{\rm c}^2} = -\frac{1}{2}.
\ee
The superconducting state has a lower free energy than the normal state for $0\leq t < 1$, demonstrating that  the superconducting state is the ground state in this temperature interval.  

\begin{figure}
\includegraphics[width=3.2in]{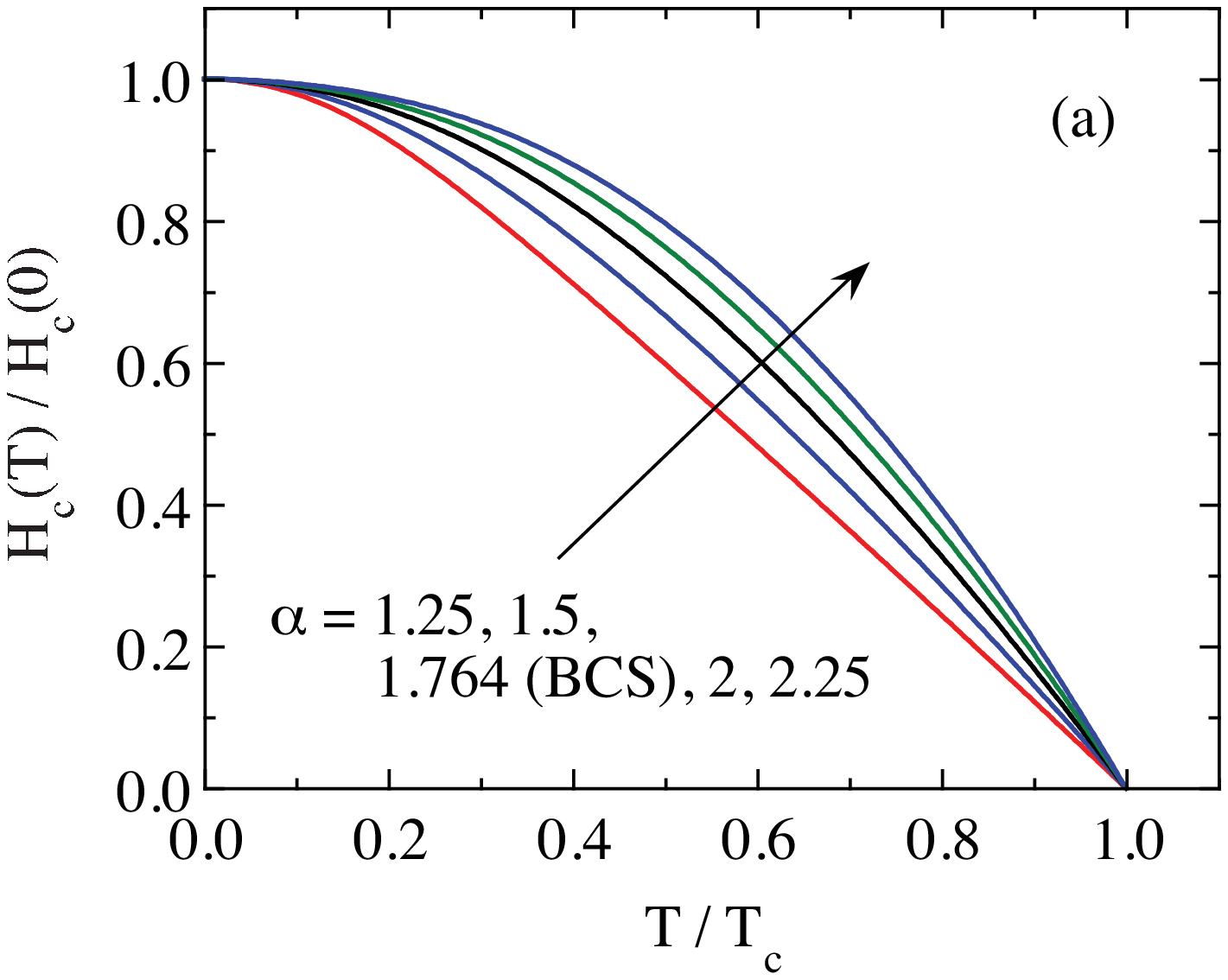}
\includegraphics[width=3.2in]{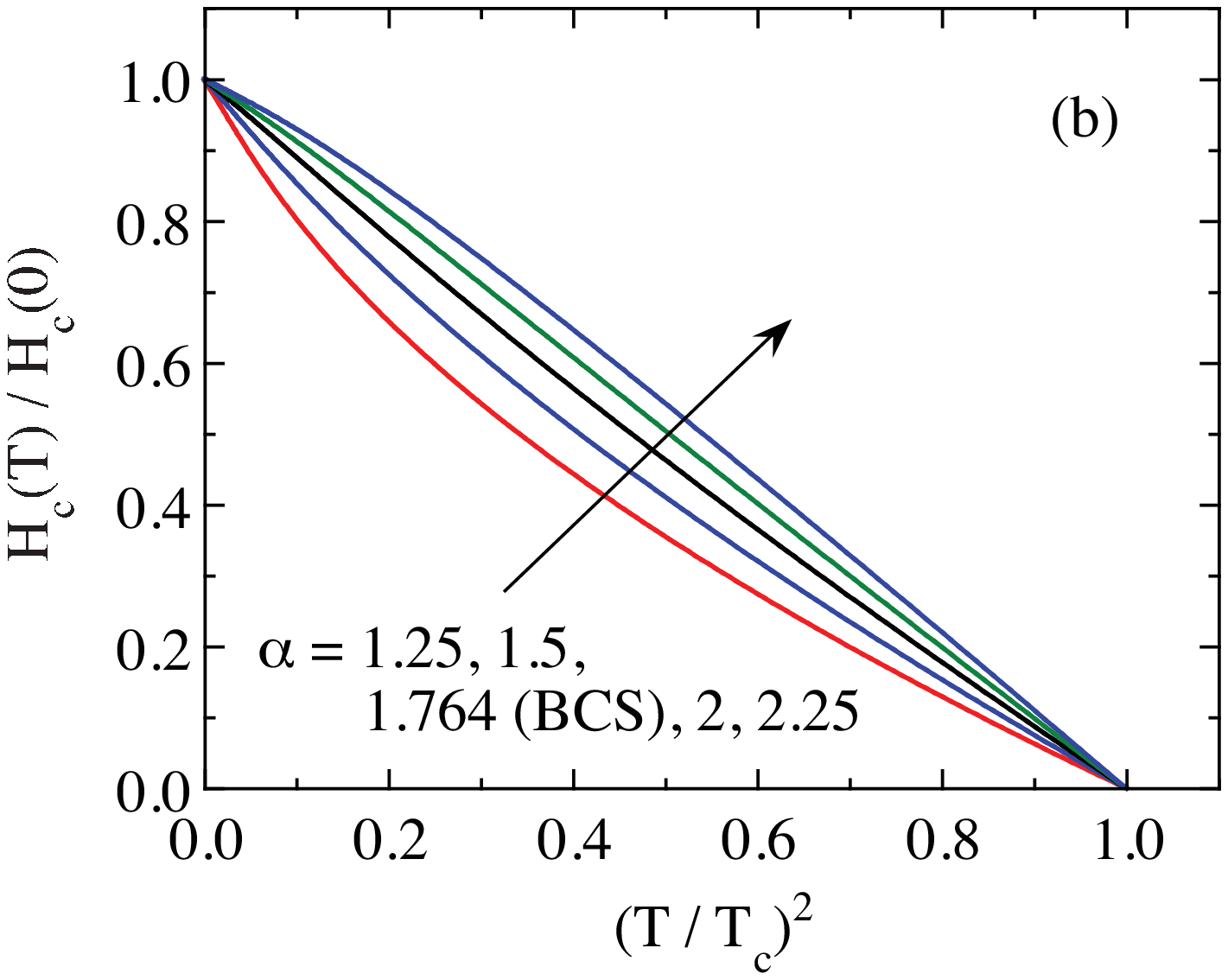}
\caption{(Color online) Thermodynamic critical field $H_{\rm c}$ calculated using Eqs.~(\ref{Eq:Fen}), (\ref{Eq:FesFen}) and (\ref{Eq:DeltaFe}), normalized by its zero-temperature value $H_{\rm c}(0)$ in Eq.~(\ref{Eq:Hc0}), (a) versus reduced temperature $t=T/T_{\rm c}$ and (b) versus $t^2$, according to the $\alpha$-model for the $\alpha$ values listed, including the BCS value $\alpha = \alpha_{\rm BCS}$ in Eq.~(\ref{Eq:alphaBCS}).}
\label{Fig:Alpha_model_HcOnHc0} 
\end{figure}

The thermodynamic critical field $H_{\rm c}(T)$ is defined in terms of the free energy difference at zero applied magnetic field between the normal and superconducting states versus temperature as
\be
\frac{H_{\rm c}^2(t)}{\gamma_{\rm n}T_{\rm c}^2} = 8\pi\frac{F_{\rm en}(t)-F_{\rm es}(t)}{\gamma_{\rm n}T_{\rm c}^2}. 
\label{Eq:DeltaFe}
\ee
Since $H_{\rm c}$ is expressed in cgs units of Oe, where ${\rm 1~Oe^2 = 1~erg/cm^3}$, the free energy is normalized to unit volume with units of ${\rm erg/cm^3}$ and $\gamma_{\rm n}$ is expressed in units of ${\rm erg/cm^3\,K^2}$.  The value of  $H_{\rm c}(T=0)$ obtained from Eqs.~(\ref{Eq:Fen}), (\ref{Eq:FesFen}) and~(\ref{Eq:DeltaFe}) is 
\be
\frac{H_{\rm c}(0)}{\left(\gamma_{\rm n}T_{\rm c}^2\right)^{1/2}} = \sqrt{\frac{6}{\pi}}\ \alpha_{\rm BCS} = \sqrt{6\pi}\,e^{-\gamma_{\rm E}}\approx 2.438,
\label{Eq:Hc0}
\ee
where the expression for $\alpha_{\rm BCS}$ in Eq.~(\ref{Eq:alphaBCS}) was used to obtain the second equality.  A plot of $H_{\rm c}/H_{\rm c}(0)$ versus $t$ calculated numerically using Eqs.~(\ref{Eq:Fen}), (\ref{Eq:FesFen}) and~(\ref{Eq:DeltaFe}) for $\alpha=\alpha_{\rm BCS}$ is shown in Fig.~\ref{Fig:Alpha_model_HcOnHc0}(a) and versus $t^2$ in Fig.~\ref{Fig:Alpha_model_HcOnHc0}(b).  From the latter figure one sees that $H_{\rm c}(t)/H_{\rm c}(0)\approx 1-t^2$ as noted by BCS\@.  A list of $H_{\rm c}(t)/H_{\rm c}(0)$ values versus $t$ for $\alpha=\alpha_{\rm BCS}$ is given in Table~\ref{Tab:AlphaModelHcH0} in the Appendix.\cite{MuhlError}

One can also determine $H_{\rm c}(t)$ from the difference between the normal- and superconducting-state entropies $S_{\rm en}(t) - S_{\rm es}(t)$ according to\cite{Reif1965}
\bea
\frac{H_{\rm c}^2}{\gamma_{\rm n}T_{\rm c}^2} &=& 8\pi \int_t^1 \left[\frac{S_{\rm en}(t^\prime)}{\gamma_{\rm n}T_{\rm c}}- \frac{S_{\rm es}(t^\prime)}{\gamma_{\rm n}T_{\rm c}}\right]dt^\prime\nonumber\\*
&=& 8\pi \int_t^1 \left[t^\prime- \frac{S_{\rm es}(t^\prime)}{\gamma_{\rm n}T_{\rm c}}\right]dt^\prime,
\label{Eq:Hc2}
\eea
where $S_{\rm en}(t^\prime)/\gamma_{\rm n}T_{\rm c} = t^\prime$, and $S_{\rm es}(t^\prime)/\gamma_{\rm n}T_{\rm c}$ is given in Eq.~(\ref{Eq:Se6}).  Equation~(\ref{Eq:Hc2}) is used to extract $H_{\rm c}(T)$ from experimental heat capacity data after subtracting the lattice and possible magnetic contributions.  Within BCS theory, we find that the normalized $H_{\rm c}/\sqrt{\gamma_{\rm n}T_{\rm c}^2}$ versus $t$ calculated from Eq.~(\ref{Eq:Hc2}) is identical to that calculated from the free energy in Eq.~(\ref{Eq:DeltaFe}) and plotted in Fig.~\ref{Fig:Alpha_model_HcOnHc0}(a) for $\alpha=\alpha_{\rm BCS}$, as must necessarily be the case.

\section{\label{BCSPenDepth} Magnetic Field Penetration Depth}

The magnetic field penetration depth $\lambda(T)$ of a superconductor is defined as the length scale of penetration of an external magnetic induction into a semi-infinite superconductor with the field applied parallel to the flat surface of the superconductor.\cite{Prozorov2006, Meservey1969}  Taking the external field direction as the $z$-direction and the direction perpendicular to the surface as the $x$-direction, the general definition is\cite{Bardeen1957, Tinkham1975}
\be
\lambda(T) = \frac{1}{B_{{\rm ext}\,z}}\int_0^\infty B_z(x,T)dx,
\label{Eq:lambdaFundDef}
\ee
where ${\bf H}_{\rm ext} = {\bf B}_{\rm ext}$  are the applied magnetic field and magnetic induction and {\bf B} is the magnetic induction inside the superconductor.  Two limiting regimes for $\lambda$ are discussed by BCS\@.  The London limit with $\lambda \equiv \lambda_{\rm L}$ corresponds to $\xi_0/\lambda_{\rm L} \ll 1$, where
\be
\xi_0 = \frac{\hbar v_{\rm F}}{\pi\Delta(0)} =\frac{1}{\pi\alpha_{\rm BCS}} \frac{\hbar v_{\rm F}}{k_{\rm B}T_{\rm c}} \approx 0.1805\,\frac{\hbar v_{\rm F}}{k_{\rm B}T_{\rm c}}
\ee
is the BCS coherence length, which is the minimum length scale over which $\Delta$ can change significantly  at low temperatures $t\ll1$, and $v_{\rm F}$ is the Fermi velocity~(speed).  Here local electrodynamics is   used, which apply to type-II superconductors.  Most superconducting compounds are in this regime.  The other limit is the Pippard limit $\xi_0/\lambda_{\rm P} \gg 1$ with $\lambda_{\rm P}\equiv\lambda$ in which nonlocal electrodynamics is important and which applies to extreme type-I clean superconductors such as pure Al with $T_{\rm c} = 1.2$~K.\cite{Meservey1969}  In the BCS paper, quasiparticle scattering by impurities  is not included in any of the calculations.  This ``clean limit'' corresponds to $\ell\gg \max(\xi_0,\lambda)$, where $\ell$ is the mean free path for quasiparticle scattering by impurities.  BCS commented that $\lambda$ is expected to increase with decreasing~$\ell$, as elaborated by Tinkham.\cite{Tinkham1975}

\subsection{London Local Electrodynamics}

In the London local limit of the electrodyamics where $\xi_0/\lambda \ll 1$, using the notation in Eq.~(\ref{Eq:lambdaFundDef}) the London equations yield\cite{Tinkham1975}
\be
B_z(x,T) = B_{{\rm ext}\,z}(x=0)\,e^{-x/\lambda_{\rm L}(T)},
\ee
where $\lambda_{\rm L}\equiv \lambda$ is the London penetration depth.  The superconducting current density {\bf J}({\bf r}) is related to the magnetic vector potential {\bf A}({\bf r}) in the London gauge by
\be
{\bf J}({\bf r}) = -\frac{1}{c\Lambda(t)}{\bf A}({\bf r}),
\ee
where for a free electron gas one has
\be
\Lambda(0) = \frac{m}{ne^2},
\label{Eq:Lambda0Def}
\ee
\be
\lambda_{\rm L}(0) = \sqrt{\frac{\Lambda(0)c^2}{4\pi}} = \sqrt{\frac{mc^2}{4\pi n e^2}} = \frac{c}{\omega_{\rm p}},
\ee 
\be
\omega_{\rm p} = \sqrt{\frac{4\pi n e^2}{m}},
\ee
$c$~is the speed of light in vacuum, $m$ is the electron mass, $n$ is the normal-state conduction electron density, $e$ is the fundamental electric charge and $\omega_{\rm p}$ is the plasma angular frequency of the conduction electrons in the normal state.\cite{Kittel2005} 

 The BCS solution for the London parameter $\Lambda(t)$ and the corresponding superfluid density $\rho_{\rm s}(t)$ in the two-fluid model is
\label{Eqs:Lambda}
\be
\frac{\rho_{\rm s}(t)}{\rho_{\rm s}(0)} = \frac{\Lambda(0)}{\Lambda(t)} = 1-I(t),
\label{Eq:LambdaI(t)}
\ee
where
\bse
\label{Eqs:I(t)}
\bea
I(t) &=& \frac{1}{2}\int_0^\infty{\rm sech}^2(z)dy,\label{Eq:Integralz}\\*
{\rm with} \ \ z &=& \frac{1}{2}\sqrt{y^2 + (\alpha_{\rm BCS}\tilde{\Delta}/t)^2}.
\eea
\ese
In the limit $t\to0$, one can set $\tilde{\Delta}=1$ and replace ${\rm sech}(z)$ by $2\,e^{-z}$ in the integral~(\ref{Eq:Integralz}).  Then the integral can be evaluated analytically, yielding
\be
I(t\to0) = \sqrt{\frac{2\pi\alpha_{\rm BCS}}{t}}\,e^{-\alpha_{\rm BCS}/t}.
\label{Eq:I(tto0)}
\ee

BCS relate $\Lambda(t)$ to the temperature derivative of the gap according to
\bse
\label{Eqs:LambdaDerivs}
\be
\frac{\Lambda(t)}{\Lambda(0)} = 1 - \frac{t}{\tilde{\Delta}(t)}\,\frac{d\tilde{\Delta}(t)}{dt} = 1 - \frac{t}{2\tilde{\Delta}^2(t)}\,\frac{d\tilde{\Delta}^2(t)}{dt}.
\label{Eq:lambdaratio1}
\ee
Then using Eq.~(\ref{Eq:LambdaI(t)}) one obtains
\be
\frac{t}{\tilde{\Delta}(t)}\,\frac{d\tilde{\Delta}(t)}{dt} = \frac{t}{2\tilde{\Delta}^2(t)}\,\frac{d\tilde{\Delta}^2(t)}{dt} = 1-\frac{1}{1-I(t)}.
\label{Eq:GapFromLambda}
\ee
\ese
In the limit $t\to1$, combining Eqs.~(\ref{Eq:limtto1D}) and~(\ref{Eq:GapFromLambda}) gives
\be
\frac{1}{1-I(t\to1)} = 1 + \frac{1}{2(1-t)}
\label{Eq:Itto1}
\ee 

The BCS prediction for $\lambda_{\rm L}(t)$ is, using Eq.~(\ref{Eq:LambdaI(t)}),\cite{BCSError}
\bse
\label{Eq:lambdaratio}
\be
\hspace{-0.5in}\frac{\lambda_{\rm L}(t)}{\lambda_{\rm L}(0)} = \sqrt{\frac{\Lambda(t)}{\Lambda(0)}} = \frac{1}{\sqrt{1-I(t)}},
\label{Eq:lambdaratio2}
\ee
which can be written
\be
\frac{\Delta\lambda_{\rm L}(t)}{\lambda_{\rm L}(0)}\equiv \frac{\lambda_{\rm L}(t) - \lambda_{\rm L}(0)}{\lambda_{\rm L}(0)} = \frac{1}{\sqrt{1-I(t)}} - 1.
\label{Eq:lambdaratio3}
\ee
\ese
From Eqs.~(\ref{Eq:lambdaratio1}) and~(\ref{Eq:lambdaratio2}) and Fig.~\ref{Fig:BCSGapVsT_AB}, $\lambda_{\rm L}(t)/\lambda_{\rm L}(0)$ is equal to unity at $t=0$ and diverges to~$\infty$ for $t\to1$.  BCS commented that superconductors that are not in the London limit at low temperatures would be expected to come into that limit for $t\to1$ as $\lambda$ diverges.  

At low temperatures $t\ll1$ one has $I(t)\ll 1$ from Eq.~(\ref{Eq:I(tto0)}), so one can Taylor expand Eq.~(\ref{Eq:lambdaratio2}) about $I(t)=0$ and use Eq.~(\ref{Eq:I(tto0)}) to obtain the low-$t$ approximations\cite{Prozorov2006}
\bse
\be
\frac{\lambda_{\rm L}(t\to0)}{\lambda_{\rm L}(0)} =1 + \frac{1}{2}I(t)= 1 + \sqrt{\frac{\pi\alpha_{\rm BCS}}{2t}}\ e^{-\alpha_{\rm BCS}/t}
\label{Eq:lambdaratioLowT}
\ee
and
\be
\frac{\Delta\lambda_{\rm L}(t\to0)}{\lambda_{\rm L}(0)} =\sqrt{\frac{\pi\alpha_{\rm BCS}}{2t}}\ e^{-\alpha_{\rm BCS}/t}.
\label{Eq:DlambdaLowT}
\ee
For $t\to1$, inserting the expression in Eq.~(\ref{Eq:Itto1}) for \mbox{$[1-I(t\to1)]^{-1}$} into~(\ref{Eq:lambdaratio2}) gives
\be
\frac{\lambda_{\rm L}(t\to1)}{\lambda_{\rm L}(0)} = \sqrt{1+\frac{1}{2(1-t)}},
\label{Eq:lambdaHighT}
\ee
\ese
which diverges to~$\infty$ at $t=1$ as noted above.  Very close to $T_{\rm c}$, Eq.~(\ref{Eq:lambdaHighT}) becomes $\lambda_{\rm L}(t\to1)/\lambda_{\rm L}(0) \approx 1/\sqrt{2(1-t)}$.

\begin{figure}
\includegraphics[width=3.3in]{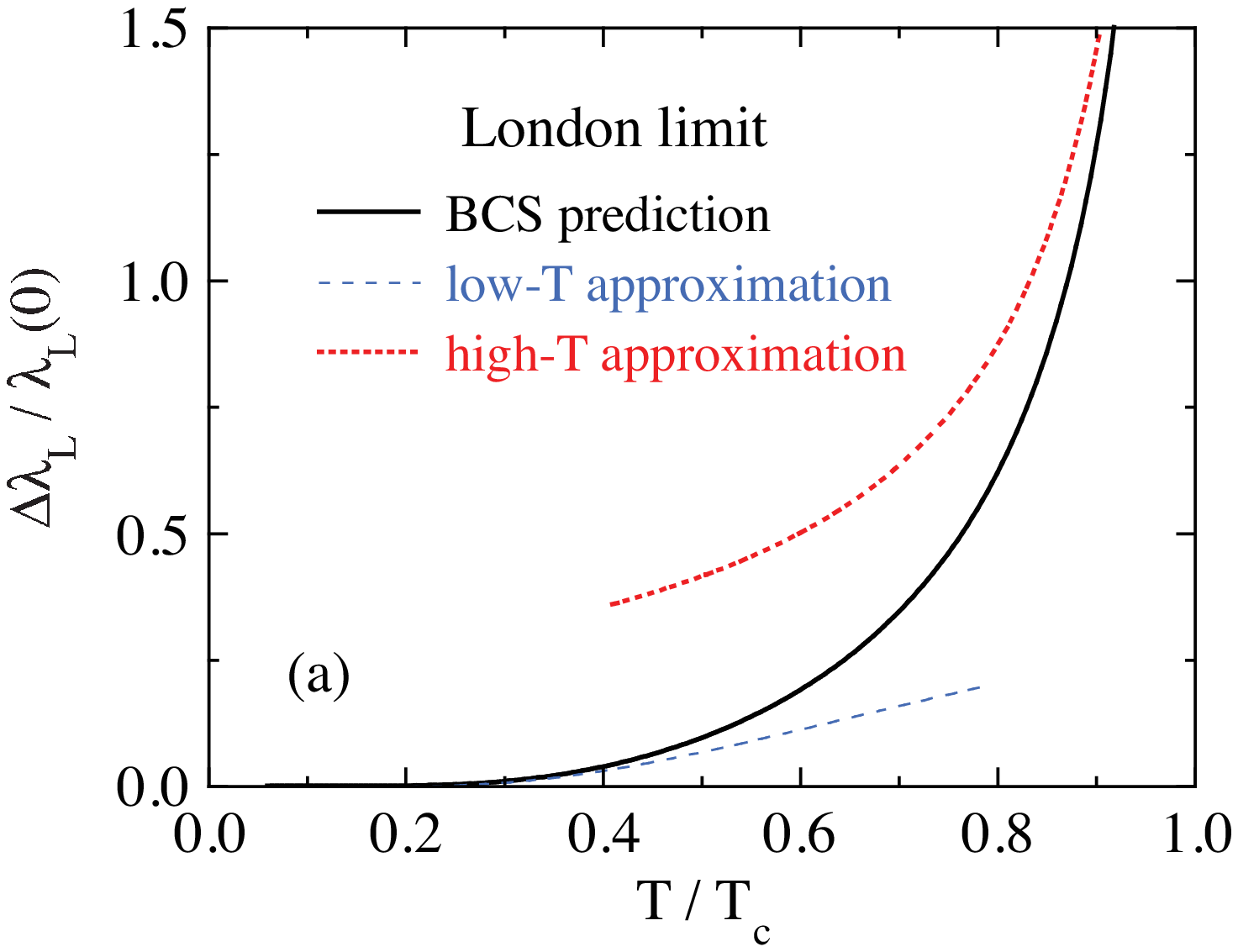}\vspace{-0.1in}
\includegraphics[width=3.4in]{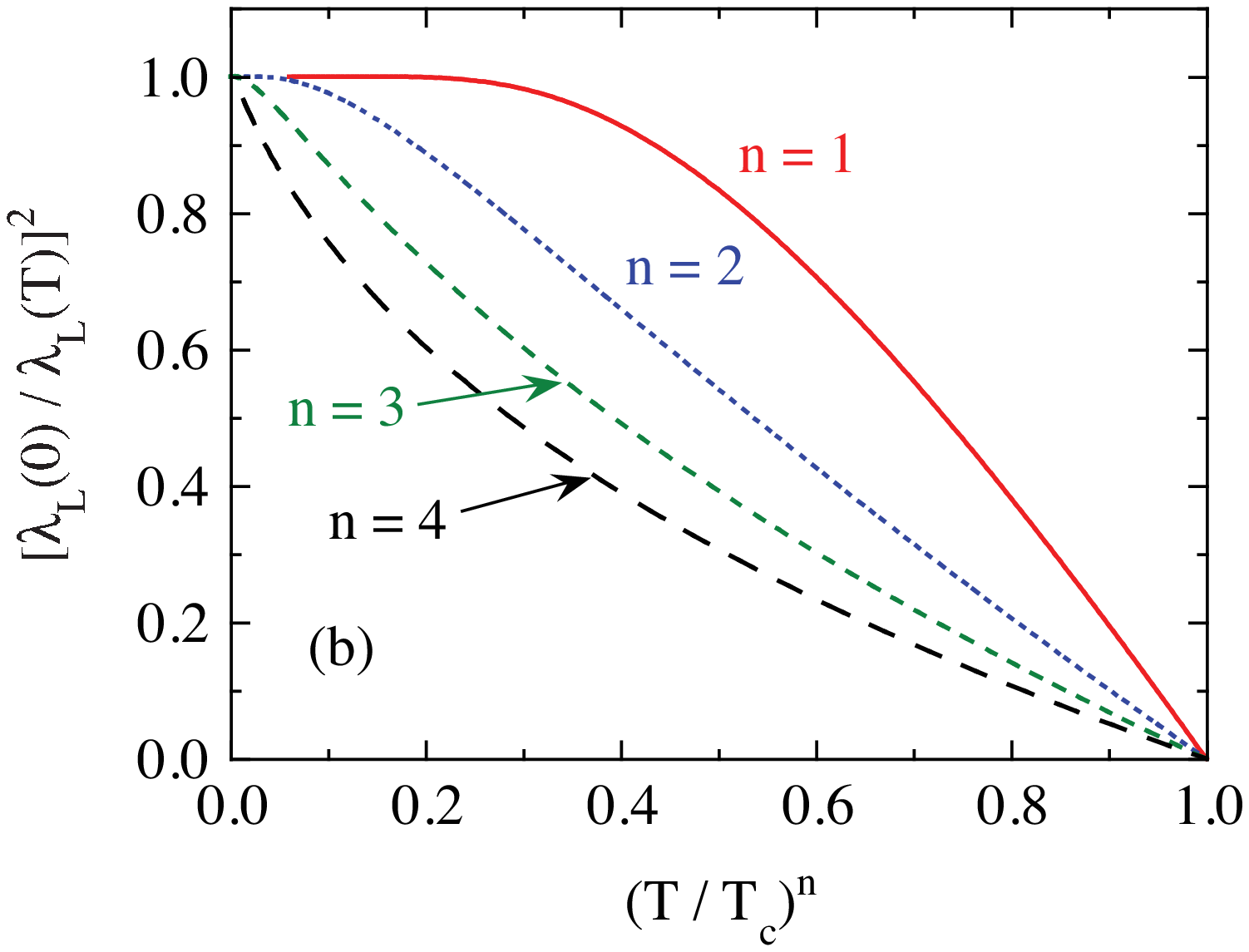}
\caption{(Color online) (a) Variation of the BCS London penetration depth $\frac{\Delta\lambda_{\rm L}(t)}{\lambda_{\rm L}(0)} \equiv \frac{\lambda_{\rm L}(t)}{\lambda_{\rm L}(0)} - 1$ versus reduced temperature $t =T/T_{\rm c}$ in the London limit $\xi_0/\lambda_{\rm L} \ll 1$ computed using Eqs.~(\ref{Eqs:I(t)}) and~(\ref{Eq:lambdaratio}). The low-$T$ approximation in Eq.~(\ref{Eq:lambdaratioLowT}) and the high-$T$ approximation in Eq.~(\ref{Eq:lambdaHighT}) are also shown. (b) Inverse penetration depth squared, $\left[\frac{\lambda_{\rm L}(0)}{\lambda_{\rm L}(t)}\right]^2$, versus $t$, $t^2$, $t^3$ and~$t^4$, as indicated.  }
\label{Fig:BCS_lambda} 
\end{figure}

\begin{figure}
\includegraphics[width=3.3in]{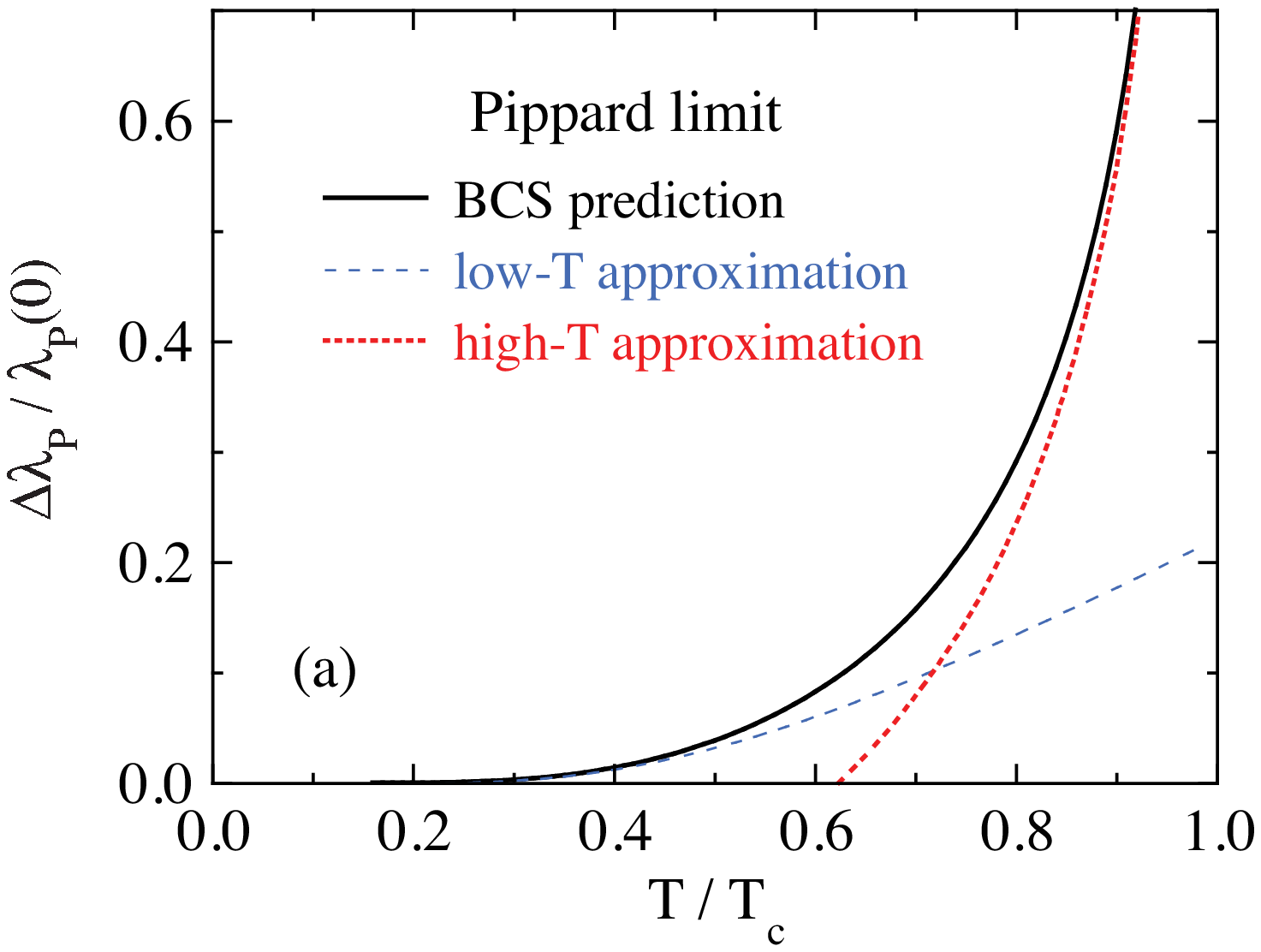}\vspace{-0.1in}
\includegraphics[width=3.4in]{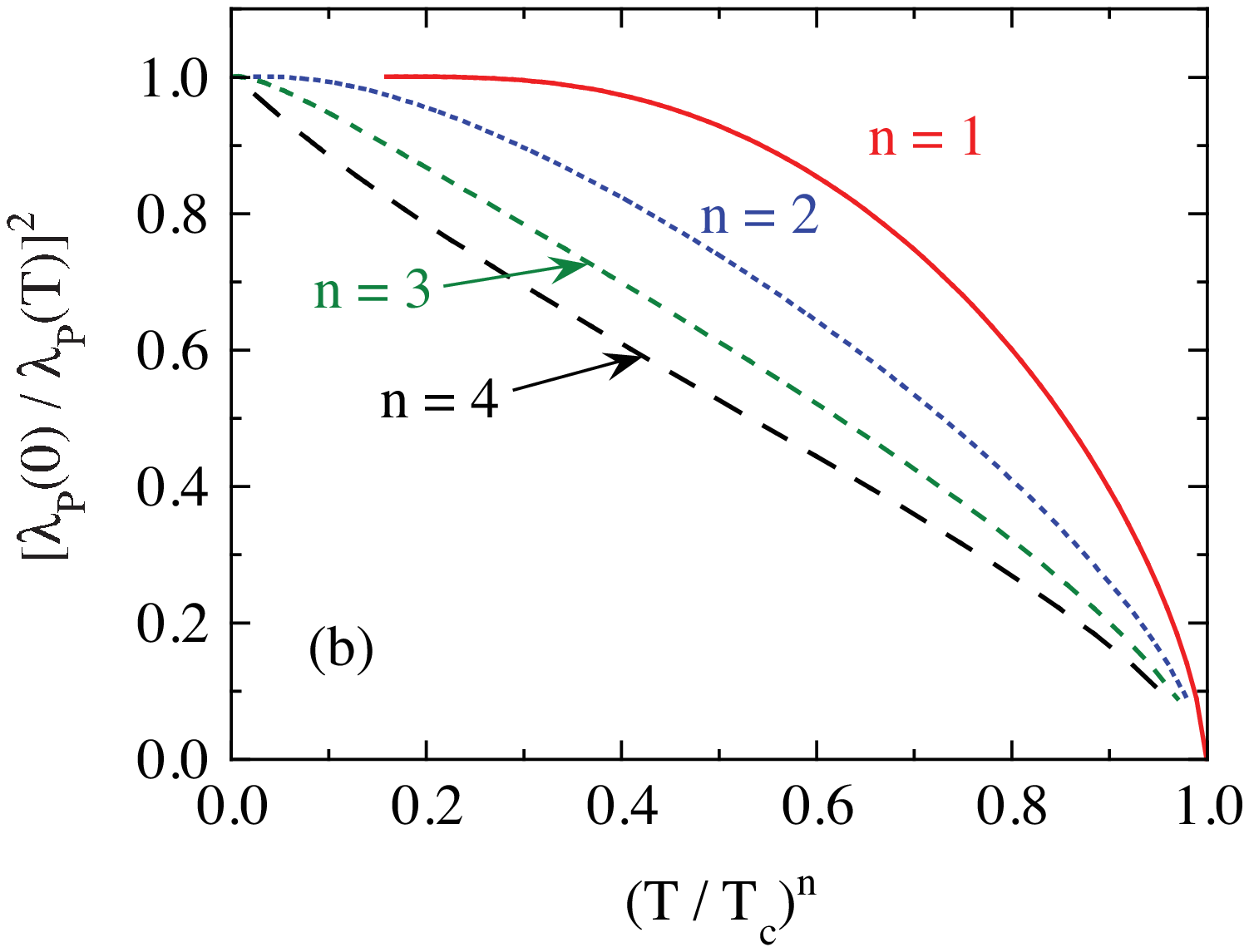}
\caption{(Color online) Plots as in Fig.~\ref{Fig:BCS_lambda}, but derived from Eqs.~(\ref{Eq:lambdaPippard})--(\ref{Eq:LambdaPippardHighT}) for the Pippard penetration depth $\lambda_{\rm P}$ in the Pippard limit $\xi_0/\lambda_{\rm P} \gg 1$.}
\label{Fig:BCS_lambda_Pippard} 
\end{figure}

A plot of $\Delta\lambda_{\rm L}(t)/\lambda_{\rm L}(0)$ versus~$t$ from Eq.~(\ref{Eq:lambdaratio3}) and a numerical solution of Eqs.~(\ref{Eqs:I(t)}) is shown in Fig.~\ref{Fig:BCS_lambda}(a), together with the low- and high-$T$ limiting behaviors in Eqs.~(\ref{Eq:lambdaratioLowT}) and~(\ref{Eq:lambdaHighT}), respectively.  A list of values of the normalized London parameter $I(t) = 1 - \frac{\Lambda(t)}{\Lambda(0)}$ from Eqs.~(\ref{Eq:LambdaI(t)}) and~(\ref{Eqs:I(t)}) and normalized London penetration depth $\frac{\Delta\lambda_{\rm L}(t)}{\lambda(0)}\equiv \frac{\lambda_{\rm L}(t)}{\lambda_{\rm L}(0)}-1 = \frac{1}{\sqrt{1-I(t)}} - 1$ from Eqs.~(\ref{Eqs:I(t)}) and (\ref{Eq:lambdaratio2}) versus~$t$ is given in Table~\ref{Tab:BCSlambda} in the Appendix.

To test power law temperature dependences, in Fig.~\ref{Fig:BCS_lambda}(b) are plotted $[\lambda_{\rm L}(0)/\lambda_{\rm L}(t)]^2$ versus~$t$, $t^2$, $t^3$ and~$t^4$.  None of the four plots are linear over the whole temperature range, but the $t^2$ dependence comes closest overall.\cite{Tinkham1975}

\subsection{Pippard Nonlocal Electrodynamics}

In the Pippard limit $\xi_0/\lambda \gg 1$ for which nonlocal electrodynamics is appropriate in clean extreme type-I superconductors, the BCS prediction for the $t$~dependence of $\lambda \equiv \lambda_{\rm P}$ for diffuse scattering of quasiparticles from the surface of the superconductor is
\bse
\be
\frac{\lambda_{\rm P}(0)}{\lambda_{\rm P}(t)} = \left\{\tilde{\Delta}(t)\tanh\left[\frac{\alpha_{\rm BCS}\tilde{\Delta}(t)}{2t}\right]\right\}^{1/3},
\label{Eq:lambdaPippard}
\ee
where 
\be
\frac{\lambda_{\rm P}(0)}{\lambda_{\rm L}(0)} = \left[\frac{\sqrt{3}\,\,\xi_0}{2\pi\lambda_{\rm L}(0)}\right]^{1/3}.
\label{Eq:lambdaP0}
\ee
\ese
Since $\xi_0/\lambda \gg 1$ in the Pippard limit, Eq.~(\ref{Eq:lambdaP0}) gives $\lambda_{\rm P}(0)>\lambda_{\rm L}(0)$, as also expected from general arguments.\cite{Tinkham1975}  A list of values of $\Delta\lambda_{\rm P}(t)/\lambda_{\rm P}(0) \equiv [\lambda_{\rm P}(t)/\lambda_{\rm P}(0)] - 1$ is given in Table~\ref{Tab:BCSlambdaPippard} in the Appendix.

For $t\to0$, we use the lowest-order expansion of $\tilde{\Delta}$ in Eq.~(\ref{Eq:DeltaLowt}) and the large argument approximation $\tanh(x) = 1-2e^{-2x}$ in Eq.~(\ref{Eq:lambdaPippard}) to obtain
\be
\frac{\lambda_{\rm P}(0)}{\lambda_{\rm P}(t\to0)} = 1 - \frac{1}{3}\left(2+\sqrt{\frac{2\pi t}{\alpha_{\rm BCS}}}\right)\,e^{-\alpha_{\rm BCS}/t} .
\label{Eq:lambdaPippardLowT}
\ee
For $t\to1$ where $\tilde{\Delta}\to0$, one can set $\tanh(x)=x$ in Eq.~(\ref{Eq:lambdaPippard}) and use Eqs.~(\ref{Eq:alphaBCS}), (\ref{Eq:dD2dtTc}) and (\ref{Eq:Delta2tto1}) to obtain
\be
\frac{\lambda_{\rm P}(0)}{\lambda_{\rm P}(t\to1)} = \left[\frac{4\pi e^{\gamma_{\rm E}}}{7\zeta(3)}(1-t)\right]^{1/3} \approx 1.3856\, (1-t)^{1/3}.
\label{Eq:LambdaPippardHighT}
\ee

A plot of $\Delta\lambda_{\rm P}(t)/\lambda_{\rm P}(0)$ versus $t$ calculated from Eq.~(\ref{Eq:lambdaPippard}) is shown in Fig.~\ref{Fig:BCS_lambda_Pippard}(a) along with the low- and high-$T$ approximations and their extrapolations using Eqs.~(\ref{Eq:lambdaPippardLowT}) and~(\ref{Eq:LambdaPippardHighT}), respectively.  Plots of $[\lambda_{\rm P}(0)/\lambda_{\rm P}(t)]^2$ versus $t^n$ ($n=1$--4) are shown in Fig.~\ref{Fig:BCS_lambda_Pippard}(b).   As noted by BCS, the calculations as displayed in Fig.~\ref{Fig:BCS_lambda_Pippard}(b) are approximately described by the Gorter-Casimir two-fluid model with $n=4$ given by
\be
\left[\frac{\lambda(0)}{\lambda(t)}\right]^2 = 1-t^4.
\label{Eq:GCmodel}
\ee

A comparison of Figs.~\ref{Fig:BCS_lambda}(a) and~\ref{Fig:BCS_lambda_Pippard}(a) shows that $\lambda_{\rm P}(t\to1)$ diverges more slowly than $\lambda_{\rm L}(t\to1)$.  However, as discussed above, a superconductor in the Pippard limit $\xi_0/\lambda \gg 1$ at low~$t$ would cross over to the London limit $\xi_0/\lambda \ll 1$ as $T_{\rm c}$ is approached due to the divergence of $\lambda$ at $T_{\rm c}$. 

\section{\label{Sec:AlphaModel} Predictions of the $\alpha$-Model}
 
The superconducting state entropy, heat capacity, free energy and thermodynamic critical field are calculated within the $\alpha$-model by replacing $\alpha_{\rm BCS}$ in Eqs.~(\ref{Eqs:SesCes}), (\ref{Eq:DCesGamTcA}), (\ref{Eq:CesLowT}), (\ref{Eq:BCSLowT}), (\ref{Eq:FesFen}) and~(\ref{Eq:Hc0}) by an adjustable parameter $\alpha\equiv \Delta(0)/k_{\rm B}T_{\rm c}$.  As noted in the introduction, the BCS $t$ dependence of the reduced gap $\tilde{\Delta}(t)$ in Fig.~\ref{Fig:BCSGapVsT_AB} calculated using Eq.~(\ref{Eq:BCSGap5}) is retained, which is carried out using the BCS value $\alpha_{\rm BCS} \approx 1.764$ in Eq.~(\ref{Eq:alphaBCS}).  From Eqs.~(\ref{Eq:lambdaratio1}) and~(\ref{Eq:lambdaratio2}), the normalized London penetration depth $\lambda_{\rm L}(t)/\lambda_{\rm L}(0)$ is uniquely related to $\tilde{\Delta}(t)$ and is hence also the same in the $\alpha$-model as in the BCS theory.

The electronic entropy $S_{\rm e}$ normalized by $\gamma_{\rm n}T_{\rm c}$ determined using Eq.~(\ref{Eq:Se6}) is plotted versus $t$ in Fig.~\ref{Fig:Alpha_Model_Ses2} for three values of $\alpha$ including the value $\alpha=\alpha_{\rm BCS}$.  One sees that the superconducting- and normal-state entropies are the same at $T_{\rm c}$ for $\alpha = 1.5$ and~2, as is the case for the BCS value $\alpha_{\rm BCS}\approx1.764$, so the transition at $T_{\rm c}$ remains second-order within the $\alpha$-model.  The electronic specific heat $C_{\rm e}/\gamma_{\rm n}T_{\rm c}$ obtained from Eq.~(\ref{Eq:Ces2}) is plotted versus $t$ in Fig.~\ref{Fig:Alpha_Model_Ces}(a) for the same three values of $\alpha$.  The specific heat jump at $T_{\rm c}$, $\Delta C_{\rm e}(T_{\rm c})/\gamma_{\rm n}T_{\rm c}$ is plotted versus $\alpha$ in Fig.~\ref{Fig:Alpha_Model_Ces}(b).  Figures~\ref{Fig:Alpha_Model_Ces}(a) and~\ref{Fig:Alpha_Model_Ces}(b) show that $\Delta C_{\rm e}(T_{\rm c})/\gamma_{\rm n}T_{\rm c}$ increases rather strongly with increasing $\alpha$. 

Since $\tilde{\Delta}(t)$ and therefore $d\tilde{\Delta}^2/dt|_{t=1}$ are assumed to be the same in the $\alpha$-model as in the BCS theory, Eqs.~(\ref{Eq:DCesGamTcA}) and~(\ref{Eq:DCesGamTc}) give the heat capacity jump at $T_{\rm c}$ as
\bea
\frac{\Delta C_{\rm e}(T_{\rm c})}{\gamma_{\rm n}T_{\rm c}} &=& \frac{\Delta C_{\rm e}(T_{\rm c})}{\gamma_{\rm n}T_{\rm c}}\bigg|_{\rm BCS}\left(\frac{\alpha}{\alpha_{\rm BCS}}\right)^2\nonumber\\*
&=& \frac{12}{7\zeta(3)}\left(\frac{\alpha}{\alpha_{\rm BCS}}\right)^2\label{Eq:DelCgamTcDelta}\\*
 &\approx&  1.426\left(\frac{\alpha}{\alpha_{\rm BCS}}\right)^2.\nonumber
\eea
The proportionality $\Delta C_{\rm e}(T_{\rm c})/\gamma_{\rm n}T_{\rm c} \propto \alpha^2$ was previously noted.\cite{Bouquet2001}  The numerical calculations of $C_{\rm es}(t)/t\gamma_{\rm n}T_{\rm c}$ at $t=1$ for the $\alpha$-model for various values of $\alpha$ in Fig.~1 of Ref.~\onlinecite{Bouquet2001} also agree with Eq.~(\ref{Eq:DelCgamTcDelta}).

The $T=0$ value $H_{\rm c}(0)$ of the thermodynamic critical field is given by Eq.~(\ref{Eq:Hc0}) as
\be
\frac{H_{\rm c}(0)}{\left(\gamma_{\rm n}T_{\rm c}^2\right)^{1/2}} = \sqrt{\frac{6}{\pi}}\ \alpha \approx 1.382\, \alpha.
\label{Eq:AlphaModelHc0}
\ee
The dependences of $H_{\rm c}/H_{\rm c}(0)$ on $t$ and $t^2$ obtained by replacing $\alpha_{\rm BCS}$ by $\alpha$ in numerical solutions of Eqs.~(\ref{Eq:FesFen}) and~(\ref{Eq:DeltaFe}) are plotted in Figs.~\ref{Fig:Alpha_model_HcOnHc0}(a) and~\ref{Fig:Alpha_model_HcOnHc0}(b), respectively, for five $\alpha$ values as listed including $\alpha_{\rm BCS}$.  On increasing $\alpha$ through $\alpha\approx 2$, the curvature of $H_{\rm c}(t)/H_{\rm c}(0)$ versus $t^2$ in Fig.~\ref{Fig:Alpha_model_HcOnHc0}(b) changes from positive to negative and the deviation from a $1-t^2$ dependence from negative to positive for $0 < t < 1$, as previously documented.\cite{Padamsee1973}

Changing $\alpha$ from the BCS value to a different value involves more than just multiplying the BCS thermodynamic results by a power of $\alpha/\alpha_{\rm BCS}$, since in addition to its presence in a prefactor, the parameter $\alpha$ is embedded within the respective integrals via the Fermi function in Eq.~(\ref{Eq:FermiFcnRed}).  Therefore, e.g., one cannot use a table of BCS thermodynamic property values versus $T$ to calculate the corresponding $T$-dependent predictions of the $\alpha$-model.  For example: (1)~This is clear from Figs.~\ref{Fig:Alpha_Model_Ses2}, \ref{Fig:Alpha_Model_Ces}(a) and~\ref{Fig:Alpha_model_HcOnHc0} where the shapes of the respective curves versus temperature strongly depend on~$\alpha$;  (2)~The superconducting and normal state entropies at $T_{\rm c}$ are not the same if one simply multiplies the BCS superconducting state $S_{\rm se}(t)/\gamma_{\rm n}T_{\rm c}$ values in Eq.~(\ref{Eq:Se6}) or Fig.~\ref{Fig:Alpha_Model_Ses2} by $(\alpha/\alpha_{\rm BCS})^2 \neq 1$, which is in conflict with our numerical calculations above which demonstrate that the $\alpha$-model predicts a second-order transition at $T_{\rm c}$; and (3)~From Eq.~(\ref{Eq:Ces2}), if one ignores the presence of $\alpha$ inside the integral, one would infer that $\Delta C_{\rm e}(T_{\rm c})/\gamma_{\rm n}T_{\rm c}\propto \alpha^3$ instead of the exact prediction with an $\alpha^2$ dependence in Eq.~(\ref{Eq:DelCgamTcDelta}).

Using Eq.~(\ref{Eq:DelCgamTcDelta}), an accurate estimate of $\alpha$ can be obtained within the $\alpha$-model from an accurate measurement of a sharp bulk heat capacity jump at $T_{\rm c}$.  However, the above discussion shows that to obtain accurate predictions of the entropy, heat capacity and thermodynamic critical field versus temperature for a given value of $\alpha$, one should do numerical calculations for that $\alpha$ using the appropriate equations.  Values of $S_{\rm es}/\gamma_{\rm n}T_{\rm c}$, $C_{\rm es}/\gamma_{\rm n}T_{\rm c}$ and $H_{\rm c}/H_{\rm c}(0)$ versus $t$ are given in the Appendix  for seven $\alpha$ values (including the BCS predictions with $\alpha = \alpha_{\rm BCS}$) in Tables~\ref{Tab:AlphaModelSes}, \ref{Tab:AlphaModelCes} and~\ref{Tab:AlphaModelHcH0}, respectively.

Perhaps surprisingly, if $\alpha_{\rm BCS}$ is consistently changed to be a fixed but arbitrary value $\alpha$ throughout the calculations, including those for $\tilde\Delta(t)$ and $d\tilde\Delta^2(t)/dt$, we find that $S_{\rm es}(t)/\gamma_{\rm n}T_{\rm c}$ and $C_{\rm es}(t)/\gamma_{\rm n}T_{\rm c}$ are independent of $\alpha$. In other words, if the $\alpha$-model is treated self-consistently, the thermodynamic properties are the same as already presented for the BCS theory in Secs.~\ref{Sec:SesCes} and~\ref{Sec:FeHc} for $\alpha=\alpha_{\rm BCS}\approx 1.764$, irrespective of the value of $\alpha$.

\section{\label{Sec:AnisOrdPars} Origins of Deviations of the Heat Capacity Jump at $T_{\rm c}$ from the BCS Prediction}

In this section we discuss superconducting gap anisotropies and other effects that can give rise to differences in  the thermodynamic properties of superconductors from the predictions of the BCS theory that the single-band $\alpha$-model is generically formulated to fit.  Thus from the fits one can quantify such deviations, which can then be interpreted in terms of other models.

If the superconducting gap is anisotropic in momentum space or scattering of the conduction carriers by magnetic impurities occurs, then one can obtain \mbox{$\alpha < \alpha_{\rm BCS}$} and $\Delta C_{\rm es}(T_{\rm c})/\gamma_{\rm n}T_{\rm c} < 1.43$ in a single-band superconductor.\cite{Mishonov2002, Openov2004}  Openov has presented a comprehensive theory on the effect of anisotropy in the superconducting order parameter on the specific heat jump $\Delta C_{\rm e}/\gamma_{\rm n}T_{\rm c}$ of BCS superconductors (i.e., within the weak-coupling mean-field approximation).\cite{Openov2004}  The calculations were carried out not only for clean superconductors as in the original BCS theory but also for those in which nonmagnetic and/or magnetic impurity scattering of the conduction electrons occurs.  We mainly discuss the clean-limit predictions here.

The wave vector {\bf k} dependent order parameter $\Delta_{\bf k}$ is defined as
\be
\Delta_{\bf k} = \Delta_0 F(\theta,\phi),
\label{Eq:FDef}
\ee
where $\Delta_0(T)$ is an angle-independent positive constant and $F$ gives its angular dependence on the Fermi surface (FS) in cylindrical [2D, $F=F(\phi)$] or spherical [3D, $F=F(\theta,\phi)$] coordinates where $\theta$ and $\phi$ are the polar and azimuthal angles, respectively.  Then the prediction is\cite{Openov2004}
\be
\frac{\Delta C_{\rm e}(T_{\rm c})}{\gamma_{\rm n}T_{\rm c}} = \left[\frac{12}{7\zeta(3)}\right]\frac{\langle\, F^2(\theta,\phi)\,\rangle^2_{\rm FS}}{\langle\, F^4(\theta,\phi)\,\rangle_{\rm FS}},
\label{Eq:DCeAnis}
\ee
where $\langle\cdots\rangle_{\rm FS}$ is an average of the enclosed quantity over the Fermi surface and the prefactor in square brackets is the heat capacity jump $\Delta C_{\rm e}(T_{\rm c})_{\rm BCS}/\gamma_{\rm n}T_{\rm c}\approx 1.426$ in Eq.~(\ref{Eq:DCesGamTc}) for a BCS superconductor with an isotropic gap ($F=1$).  
The moments of $F$ are defined as
\bea
\langle\, F^n(\phi) \,\rangle_{\rm FS} &=& \frac{1}{2\pi}\int_0^{2\pi}d\phi\,F^n(\phi)\quad ({\rm 2D}),\\*
\langle\, F^n(\theta,\phi) \,\rangle_{\rm FS} &=& \frac{1}{4\pi}\int_0^\pi d\theta\,\sin\theta \int_0^{2\pi}d\phi\,F^n(\theta,\phi) \quad ({\rm 3D}).\nonumber
\label{Eq:FMoments}
\eea
We normalize $F(\theta,\phi)$ such that $\langle\, F^2(\theta,\phi)\,\rangle_{\rm FS}=1$, and then rewrite Eq.~(\ref{Eq:DCeAnis}) as
\be
\frac{\Delta C_{\rm e}(T_{\rm c})}{\Delta C_{\rm e}(T_{\rm c})_{\rm BCS}} = \frac{1}{\langle\, F^4(\theta,\phi)\,\rangle_{\rm FS}}.
\label{Eq:DCeAnis2}
\ee

\begin{table}
\caption{Heat capacity jump at $T_{\rm c}$, $\Delta C_{\rm e}(T_{\rm c})$, normalized by the BCS value $\Delta C_{\rm e}(T_{\rm c})_{\rm BCS}$ for an isotropic $s$-wave gap in Eq.~(\ref{Eq:DCesGamTc}), and for superconductors with anisotropic gaps $\Delta = \Delta_0F(\theta,\phi)$ as in Eq.~(\ref{Eq:FDef}), where $\Delta_0$ is a positive constant and $F(\theta,\phi)$ is its angular dependence normalized such that the second moment is $\langle\, F^2(\theta,\phi)\,\rangle_{\rm FS}=1$.  The dimensionality of the FS is 2D (cylindrical) and/or 3D (spherical).}
\label{Table:Anisotropy}
\begin{ruledtabular}
\begin{tabular}{l c c}
order parameter	& $F(\theta,\phi)$ & $\Delta C_{\rm e}(T_{\rm c})/\Delta C_{\rm e}(T_{\rm c})_{\rm BCS}$ \\
\hline
2D, 3D $s$-wave				&  1					& 1 \\
2D anis. $s$-wave 			& $\frac{a + \cos^2(2\phi)}{\sqrt{(3+8a+8a^2)/8}}$	&	$\frac{2[3+8a+8a^2]^2}{35+32a(1+a)(5+4a+4a^2)}$	\\
 &($a < 1$)			&		\\
2D $d$-wave					& $\sqrt{2}\,\cos(2\phi)$	&  2/3	\\
3D axial $s$-wave 					& $\frac{1 + a\cos^2\theta}{\sqrt{1 + \frac{2a}{3} + \frac{a^2}{5}}}$		& $\frac{7(15+10a+3a^2)^2}{5(315 + 420 a + 378 a^2 + 180 a^3 + 35 a^4)}$			\\
	\end{tabular}
	\end{ruledtabular}
\end{table}

\begin{figure}
\includegraphics[width=2.4in]{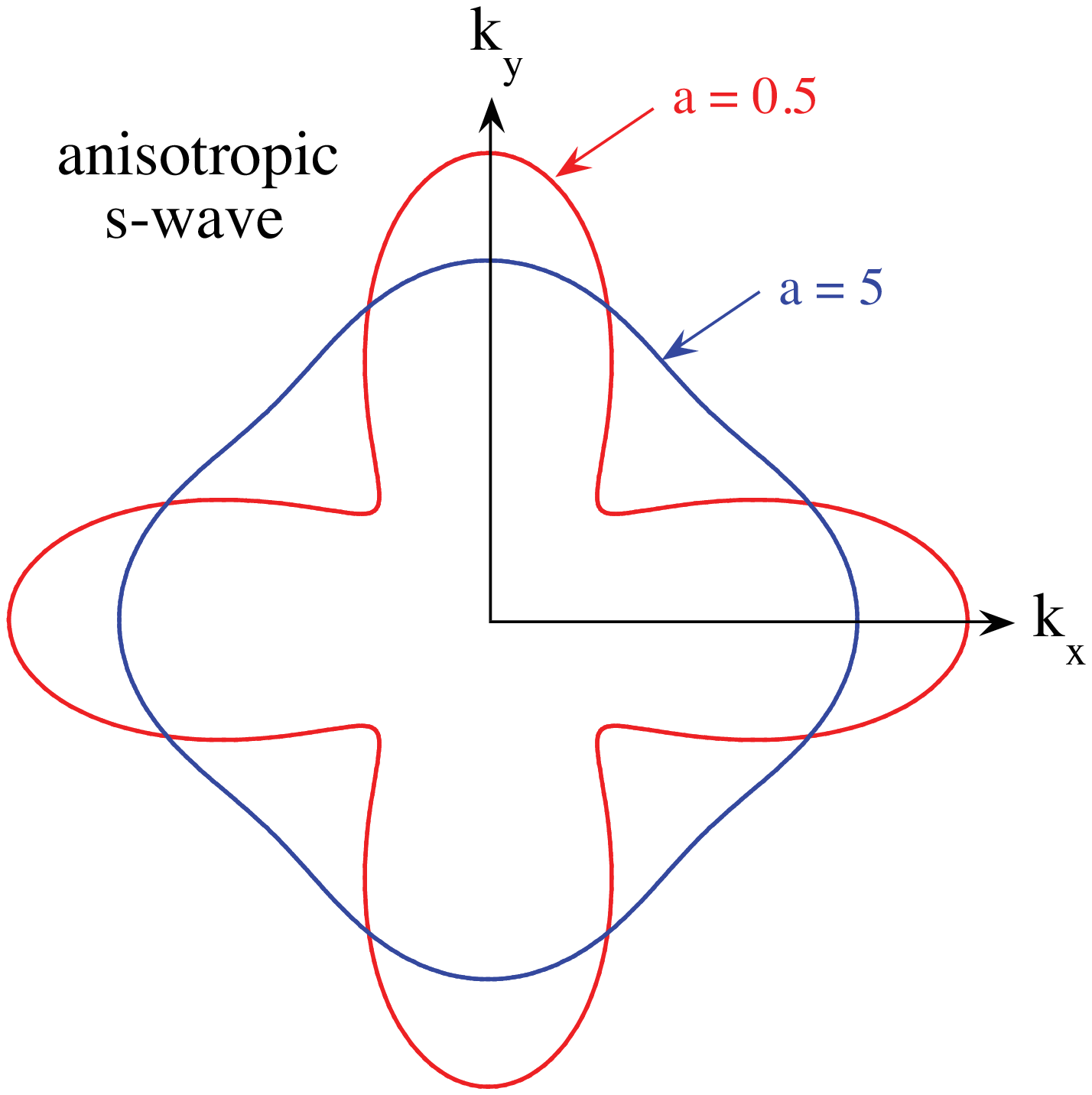}
\includegraphics[width=2.4in]{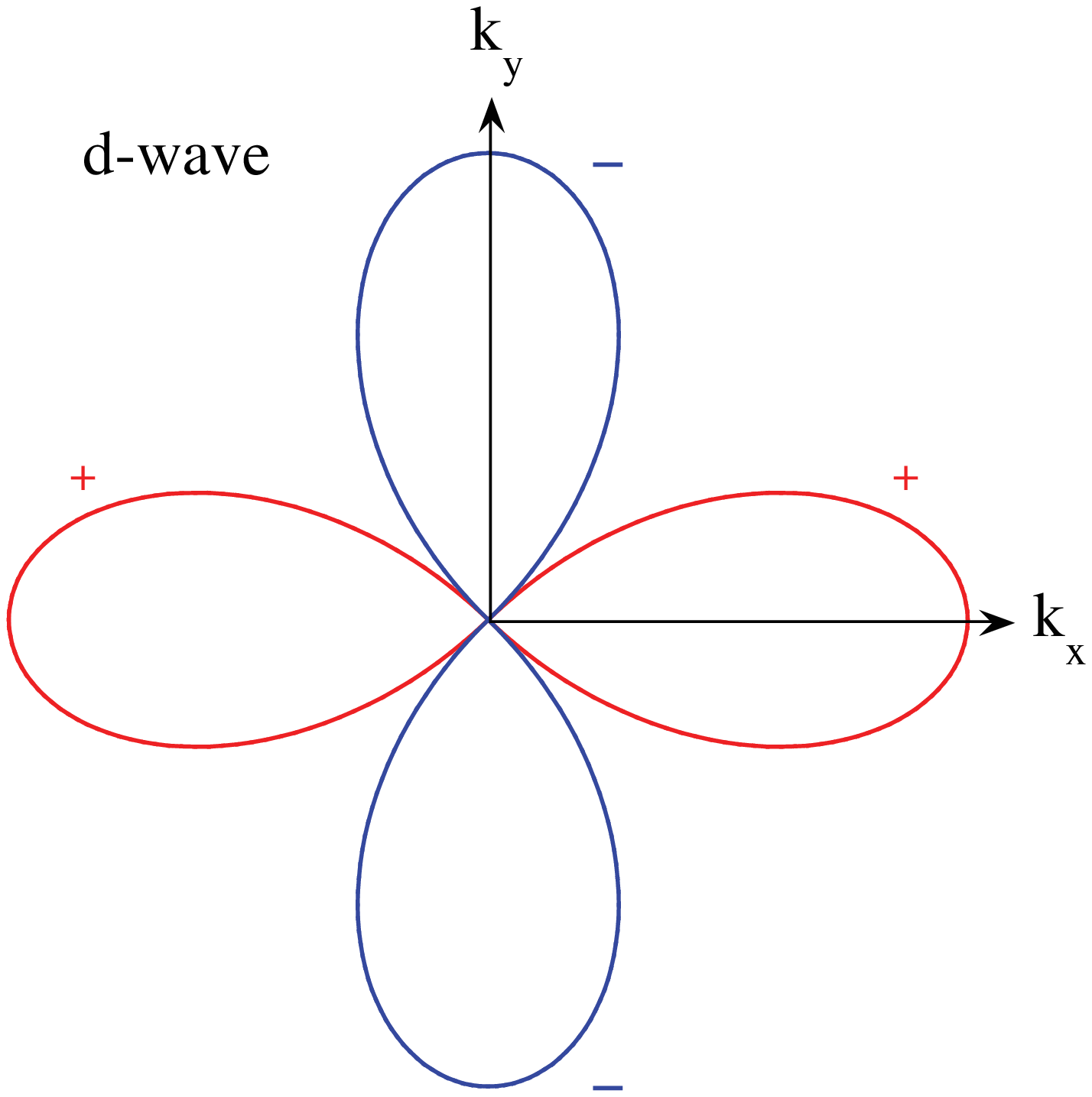}
\caption{(Color online) (top panel) Angular dependences of the 2D anisotropic $s$-wave order parameters with anisotropy parameters $a = 0.5$~(red) and $a = 0.5$~(blue)  versus azimuthal angle $\phi$ measured from the positive $k_x$-axis.  (bottom panel) 2D $d$-wave  superconducting order parameter versus $\phi$.  The anisotropic $s$-wave order parameters have the same sign on all parts of the Fermi surface, whereas the $d$-wave order parameter changes sign versus~$\phi$  as shown.  The data are computed from the expressions in Table~\ref{Table:Anisotropy}.}
\label{Fig:Anisotropic_gap} 
\end{figure}

\begin{figure}
\includegraphics[width=1.9in]{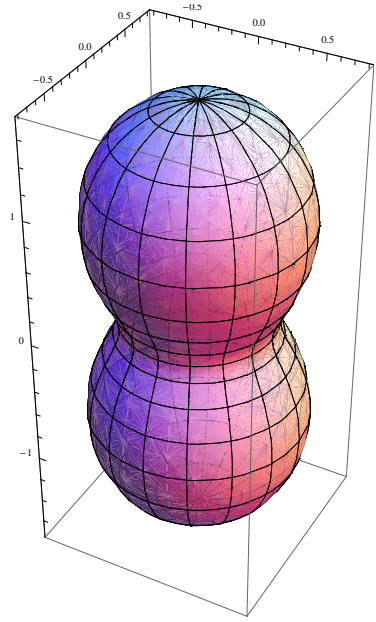}
\caption{(Color online) Angular dependence of the 3D cylindrically symmetric but axially anisotropic $s$-wave order parameter in Table~\ref{Table:Anisotropy} for anisotropy parameter $a = 2$.  The polar $z$-axis is vertical.}
\label{Fig:Anisotropic_axial_s_wave} 
\end{figure}

\begin{figure}
\includegraphics[width=3.2in]{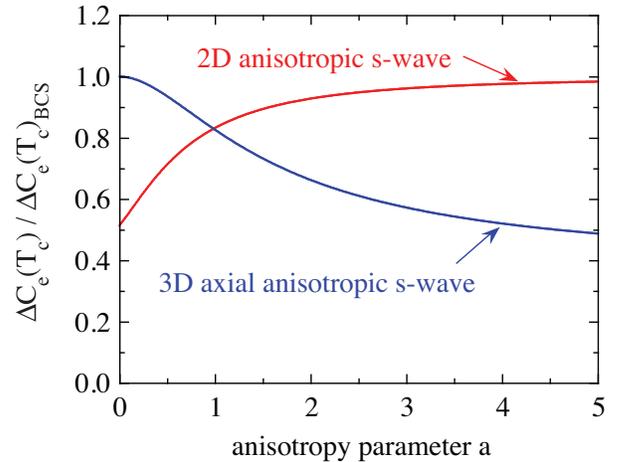}
\caption{(Color online) Heat capacity jump $\Delta C_{\rm e}(T_{\rm c})$ at $T_{\rm c}$ normalized by the BCS value $\Delta C_{\rm e}(T_{\rm c})_{\rm BCS}$ versus the anisotropy parameter $a$ in Table~\ref{Table:Anisotropy} for the 2D anisotropic $s$-wave and the 3D axial anisotropic $s$-wave order parameters.  The different behaviors versus $a$ result from the different roles of $a$ in the anisotropy expressions in Table~\ref{Table:Anisotropy}.}
\label{Fig:Ce_jump_anisotropic_gap} 
\end{figure}

For simplicity, we consider here free-electron Fermi surfaces in either two dimensions (2D, cylindrical) or three (3D, spherical).  Several types of gap anisotropy for 2D square or 3D tetragonal symmetry\cite{VanHarlingen1995} are considered as shown in Table~\ref{Table:Anisotropy}.  Polar plots of the angular dependences of the 2D anisotropic $s$-wave and $d$-wave order parameters are shown in Fig.~\ref{Fig:Anisotropic_gap}. Conventional superconductors driven by the electron-phonon interaction have $s$-wave order parameters with the same sign everywhere on the Fermi surface, whereas some unconventional superconductors such as the high-$T_{\rm c}$ cuprates have sign-changing $d$-wave order parameters.\cite{VanHarlingen1995}  An example of a 3D axial anisotropic $s$-wave gap\cite{Haas2001} is shown in Fig.~\ref{Fig:Anisotropic_axial_s_wave}.

The heat capacity jump at $T_{\rm c}$ in these cases is calculated using Eq.~(\ref{Eq:DCeAnis2}).  For 2D $d$-wave pairing, $\Delta C_{\rm e}(T_{\rm c})/\Delta C_{\rm e}(T_{\rm c})_{\rm BCS}$  has the single value of 2/3 (Table~\ref{Table:Anisotropy}).  Plots of $\Delta C_{\rm e}(T_{\rm c})/\Delta C_{\rm e}(T_{\rm c})_{\rm BCS}$ versus the respectively-defined anisotropy parameter $a$ for 2D anisotropic $s$-wave and 3D axial anisotropic $s$-wave order parameters calculated from the expressions in Table~\ref{Table:Anisotropy} using Eq.~(\ref{Eq:DCeAnis2}) are shown in Fig.~\ref{Fig:Ce_jump_anisotropic_gap}.  One sees that the heat capacity jump at $T_{\rm c}$ monotonically decreases from the BCS value with increasing anisotropy, which corresponds to decreasing $a$ for 2D anisotropic $s$-wave and increasing $a$ for 3D axial anisotropic $s$-wave order parameters.  The maximum decrease in the limit of large anisotropy is about a factor of two for both types of anisotropy.

As mentioned in the introduction, values of the specific heat jump $\Delta C_{\rm e}(T_{\rm c})/\gamma_{\rm n}T_{\rm c}$ larger than the BCS value of 1.43 are observed in moderate- and strong-coupling superconductors for which $\alpha > \alpha_{\rm BCS}$.\cite{Carbotte1990}  Combescot concluded that $\alpha$ has the upper limit $\alpha\leq 2$ within the weak-coupling BCS theory.\cite{Combescot1988}    When magnetic impurities are present in a BCS superconductor, both $T_{\rm c}$ and the normalized specific heat jump $\Delta C_{\rm e}(T_{\rm c})/\gamma_{\rm n}T_{\rm c}$ are reduced compared to the BCS values with increasing electron-impurity scattering rate.\cite{Openov2004, Skalski1964}  Both quantities go to zero at sufficiently high scattering rates.  Nonmagnetic impurites can also depress both $T_{\rm c}$ and $\Delta C_{\rm e}(T_{\rm c})/\gamma_{\rm n}T_{\rm c}$ for single-band materials with anisotropic superconducting order parameters.\cite{Openov2004}

\section{\label{Sec:Extensions} Extensions of the Single-Band $\alpha$-Model}

The single-band $\alpha$-model in the clean limit has been extended to the two-band $\alpha$-model in which each of two electron bands develop distinct isotropic superconducting energy gaps on the two respective Fermi surfaces.  The parameter $\alpha=\Delta(0)/k_{\rm B}T_{\rm c}$ and $\Delta(0)$ are different for the two bands but $T_{\rm c}$ is the same.  Within the weak-coupling theory, Eq.~(\ref{Eq:DCeAnis}) for the heat capacity jump at $T_c$ for an anisotropic single gap is generalized to\cite{Mishonov2005}
\be
\frac{\Delta C_{\rm e}(T_{\rm c})}{\gamma_{\rm n}T_{\rm c}} = \left[\frac{12}{7\zeta(3)}\right]\frac{[c_1\langle\, F_1^2(\theta,\phi)\,\rangle_{\rm FS} + c_2 \langle\, F_2^2(\theta,\phi)\,\rangle_{\rm FS}]^2}{c_1\langle\, F_1^4(\theta,\phi)\,\rangle_{\rm FS} + c_2\langle\, F_2^4(\theta,\phi)\,\rangle_{\rm FS}},
\label{Eq:DCeAnis3}
\ee
where $c_1$ and $c_2$ are the respective fractions of the total $N(0)$ contributed by bands~1 and~2, with $c_1+c_2=1$, and the prefactor in square brackets is the BCS value of $\Delta C_{\rm e}(T_{\rm c})/\gamma_{\rm n}T_{\rm c}$ in Eq.~(\ref{Eq:DCesGamTc}).

The isotropic $s$-wave two-band (two-gap) $\alpha$-model has been applied to understand the temperature dependences of the specific heat and London penetration depth of materials such as the fiducial two-gap compound MgB$_2$ with $T_{\rm c}=39$~K (Refs.~\onlinecite{Bouquet2001}, \onlinecite{Nagamatsu2001}, \onlinecite{Fletcher2005}) and ${\rm NbSe_2}$ ($T_{\rm c}=7.1$~K).\cite{Fletcher2007}  It seems likely that many previously-studied superconductors in the clean limit have multi-gap behaviors in addition to those already identified.  Dolgov et al.\ made a detailed comparison of the predictions of the two-band $\alpha$-model with the corresponding solution to the full Eliashberg equations and found good agreement.\cite{Dolgov2005}  When strong interband scattering by nonmagnetic impurities is present, within the Eliashberg formalism the two gaps merge into a single gap.\cite{Brinkman2002}  Interband scattering by \emph{nonmagnetic} impurities has a similar strong effect on the superconductivity in a two-band, two-gap superconductor as does intraband scattering by \emph{magnetic} impurities in a single-gap superconductor. 

As noted above, the $\alpha$-model is not self-consistent because  $\alpha = \alpha_{\rm BCS}$ is used to calculate $\tilde{\Delta}(t)$ and the London penetration depth, but a different $\alpha$~value is used to calculate the thermodynamic properties.  Kogan, Martin and Prozorov formulated a two-band (two-gap) weak-coupling clean-limit model with isotropic $s$-wave gaps on each Fermi surface, termed the $\gamma$-model, that self-consistently treats the temperature dependences of the two gaps and the associated London penetration depth and thermodymic properties.\cite{Kogan2009}  They find that $\Delta C_{\rm e}(T_{\rm c})/k_{\rm B}T_{\rm c}$ has an upper limit given by the BCS value of~1.43, and can be strongly suppressed from that value depending on the relative values of $N(0)$ of the two bands and the intraband and interband electron-phonon coupling constants.  The same effect was documented in Sec.~\ref{Sec:AnisOrdPars} above for anisotropy of the gap within the single-band model.  Thus when considering the effects of gap anisotropy on the heat capacity jump at $T_{\rm c}$, one should evidently consider gap anisotropy globally within the Brillouin zone.  That is, when two Fermi surfaces each have isotropic gaps, but where the gaps are different on each Fermi surface, then the gap is wave-vector-dependent within the Brillouin zone, leading to a suppression of the heat capacity jump from the BCS value qualitatively similar to that in the anisotropic single-band model.  Kogan, Martin and Prozorov applied their $\gamma$-model to precisely fit the temperature dependences of the London penetration depth and electronic heat capacity of ${\rm MgB_2}$ and ${\rm V_3Si}$ ($T_{\rm c}=17$~K).  Prozorov and Kogan have also reviewed applications of the same model to fit the temperature dependences of the London penetration depths of multi-gap FeAs-based high-$T_{\rm c}$ superconductors.\cite{Prozorov2011}

\section{\label{Summary} Summary}

The single-band $\alpha$-model has been extensively used by experimentalists in the past to fit electronic heat capacity versus temperature data for superconductors that deviate from the BCS prediction.  The model is based on the BCS theory and assumes the same values of the normalized gap and London penetration depth as predicted by BCS that are obtained using the BCS value $\alpha_{\rm BCS} \approx 1.764$ of $\alpha\equiv \Delta(0)/k_{\rm B}T_{\rm c}$.  However, in calculating the electronic entropy, heat capacity, free energy and thermodynamic critical field of the superconducting state, the $\alpha$-model takes $\alpha$ to be a variable, which then allows calculations of these thermodynamic quantities to be adjusted to fit experimental data.  This is an inconsistency in the model.

Most previous experimental papers fitting the $\alpha$ model to experimental thermodynamics data do not explain how the theoretical values used in the fit were obtained.  We have written the equations for the thermodynamic predictions of the BCS theory in terms of $\alpha$ to clarify how to calculate these quantities versus $\alpha$ and~$T$ and have compared the results for different $\alpha$ with the BCS predictions.  Tables of values of the BCS predictions of the superconducting order parameter, London parameter and London penetration depth versus temperature are given in the Appendix, which are the same in the $\alpha$-model by assumption, and of the Pippard penetration depth versus temperature.  Additional tables in the Appendix give the temperature dependences of the superconducting state electronic entropy, heat capacity and thermodynamic critical field for seven representative values of $\alpha$ including $\alpha_{\rm BCS}$ to facilitate fitting of experimental data by the $\alpha$-model.  These results could be interpolated to obtain values for other values of $\alpha$.  These tables supplement the table in Ref.~\onlinecite{Muhlschlegel1959} of superconducting state properties versus temperature predicted by the BCS theory.

We find that if the $\alpha$-model is treated self-consistently, i.e., if the same value of $\alpha$ is used to calculate the temperature dependence of the superconducting gap and of the thermodynamic properties, then the normalized thermodynamic properties versus temperature are independent of $\alpha$ and are therefore the same as presented for the BCS theory in Secs.~\ref{Sec:SesCes} and~\ref{Sec:FeHc} and Tables~\ref{Tab:AlphaModelSes}--\ref{Tab:AlphaModelHcH0} in the Appendix for $\alpha=\alpha_{\rm BCS}\approx 1.764$.

Mechanisms for producing deviations of the superconducting state thermodynamic properties from the BCS predictions were discussed that can be quantified using the $\alpha$-model.  It is well-known that strong electron-phonon coupling \emph{increases} the heat capacity jump $\Delta C_{\rm e}(T_{\rm c})/\gamma_{\rm n}T_{\rm c}$ at $T_{\rm c}$ compared to the BCS value of 1.43.\cite{Carbotte1990}  We calculated the influence of superconducting gap anisotropy in momentum space on the specific heat jump at $T_{\rm c}$ for several types of gap anisotropy using the formalism discussed in Ref.~\onlinecite{Openov2004}, and found that the jump monotonically \emph{decreases} with increasing anisotropy from the BCS value by up to a factor of two. Extensions of the $\alpha$ model were also discussed, including the two-band $\alpha$-model\cite{Mishonov2005} and the self-consistent two-band $\gamma$-model.\cite{Kogan2009, Prozorov2011}

The present work points the way towards a uniform application of the $\alpha$-model by experimentalists in their analyses of superconducting state thermodynamic data.

\acknowledgments

The author is grateful to V. K. Anand, R. M. Fernandes, V. G. Kogan and R. Prozorov for helpful discussions and correspondence.  This research was supported by the U.S. Department of Energy, Office of Basic Energy Sciences, Division of Materials Sciences and Engineering.  Ames Laboratory is operated for the U.S. Department of Energy by Iowa State University under Contract No.~DE-AC02-07CH11358.

\appendix*
\section{Tables of Values}

Table~\ref{Tab:BCSGap} gives the $T$ dependence of the BCS superconducting gap with high resolution in $t$, especially near $T_{\rm c}$, a dependence that is retained in the $\alpha$-model.  The subsequent three tables give the $T$ dependences of the electronic entropy (Table~\ref{Tab:AlphaModelSes}), electronic heat capacity (Table~\ref{Tab:AlphaModelCes}) and thermodynamic critical field \mbox{(Table~\ref{Tab:AlphaModelHcH0})} for seven values of $\alpha$ including the BCS value $\alpha_{\rm BCS}\approx 1.764$, where the exact value of $\alpha_{\rm BCS}$ in Eq.~(\ref{Eq:alphaBCS}) was used in computing the values in the tables. Values of the BCS London parameter and the London penetration depth versus $t$ are given in Table~\ref{Tab:BCSlambda} and values of the Pippard penetration depth versus $t$ are given in Table~\ref{Tab:BCSlambdaPippard}. Table~\ref{Tab:BCSlambda} also applies to the $\alpha$-model.

%\clearpage
%\vspace{6in}

\begin{table*}
\caption{Normalized superconducting order parameter $\tilde{\Delta}(t)\equiv \Delta(t)/\Delta(0)$ versus reduced temperature $t = T/T_{\rm c}$ calculated within the BCS weak-coupling theory using Eq.~(\ref{Eq:BCSGap5}).    The data are the same in the $\alpha$-model.}
\label{Tab:BCSGap}
\begin{ruledtabular}
\begin{tabular}{c c c c c c}
$t$ & $\tilde{\Delta}(t)$ & $t$ & $\tilde{\Delta}(t)$ & $t$ & $\tilde{\Delta}(t)$ \\
\hline
0 & 1 & 0.760000 & 0.76399 & 0.997246 & 0.091038 \\
0.020000 & 1.00000 & 0.780000 & 0.73863 & 0.997545 & 0.085956 \\
0.022763 & 1.00000 & 0.781224 & 0.73701 & 0.997812 & 0.081157 \\
0.040000 & 1.00000 & 0.800000 & 0.71104 & 0.998050 & 0.076625 \\
0.060000 & 1.00000 & 0.805016 & 0.70374 & 0.998262 & 0.072345 \\
0.080000 & 1.00000 & 0.820000 & 0.68095 & 0.998451 & 0.068303 \\
0.100000 & 1.00000 & 0.826220 & 0.67103 & 0.998620 & 0.064487 \\
0.120000 & 1.00000 & 0.840000 & 0.64801 & 0.998770 & 0.060883 \\
0.129036 & 1.00000 & 0.845118 & 0.63906 & 0.998904 & 0.057480 \\
0.140000 & 1.00000 & 0.860000 & 0.61173 & 0.999023 & 0.054268 \\
0.160000 & 0.99999 & 0.861962 & 0.60797 & 0.999129 & 0.051234 \\
0.180000 & 0.99996 & 0.876973 & 0.57786 & 0.999224 & 0.048370 \\
0.200000 & 0.99988 & 0.880000 & 0.57148 & 0.999308 & 0.045666 \\
0.220000 & 0.99971 & 0.890352 & 0.54880 & 0.999383 & 0.043113 \\
0.223753 & 0.99967 & 0.900000 & 0.52634 & 0.999450 & 0.040702 \\
0.240000 & 0.99941 & 0.902276 & 0.52084 & 0.999510 & 0.038426 \\
0.260000 & 0.99892 & 0.912904 & 0.49400 & 0.999563 & 0.036277 \\
0.280000 & 0.99818 & 0.920000 & 0.47491 & 0.999611 & 0.034249 \\
0.300000 & 0.99712 & 0.922375 & 0.46829 & 0.999653 & 0.032334 \\
0.308169 & 0.99659 & 0.930817 & 0.44371 & 0.999691 & 0.030525 \\
0.320000 & 0.99569 & 0.938340 & 0.42024 & 0.999725 & 0.028818 \\
0.340000 & 0.99382 & 0.940000 & 0.41484 & 0.999755 & 0.027206 \\
0.360000 & 0.99146 & 0.945046 & 0.39787 & 0.999781 & 0.025685 \\
0.380000 & 0.98854 & 0.951022 & 0.37657 & 0.999805 & 0.024248 \\
0.383405 & 0.98799 & 0.956348 & 0.35631 & 0.999826 & 0.022892 \\
0.400000 & 0.98503 & 0.960000 & 0.34161 & 0.999845 & 0.021612 \\
0.420000 & 0.98088 & 0.961095 & 0.33705 & 0.999862 & 0.020403 \\
0.440000 & 0.97603 & 0.965326 & 0.31877 & 0.999877 & 0.019262 \\
0.450459 & 0.97320 & 0.969097 & 0.30141 & 0.999890 & 0.018184 \\
0.460000 & 0.97044 & 0.972458 & 0.28495 & 0.999902 & 0.017167 \\
0.480000 & 0.96407 & 0.975453 & 0.26935 & 0.999913 & 0.016207 \\
0.500000 & 0.95688 & 0.978122 & 0.25456 & 0.999922 & 0.015300 \\
0.510221 & 0.95288 & 0.980000 & 0.24358 & 0.999931 & 0.014444 \\
0.520000 & 0.94883 & 0.980502 & 0.24056 & 0.999938 & 0.013637 \\
0.540000 & 0.93986 & 0.982622 & 0.22730 & 0.999945 & 0.012874 \\
0.560000 & 0.92993 & 0.984512 & 0.21476 & 0.999951 & 0.012154 \\
0.563484 & 0.92810 & 0.986196 & 0.20288 & 0.999956 & 0.011474 \\
0.580000 & 0.91899 & 0.987697 & 0.19165 & 0.999961 & 0.010832 \\
0.600000 & 0.90699 & 0.989035 & 0.18103 & 0.999965 & 0.010226 \\
0.610955 & 0.89995 & 0.990228 & 0.17099 & 0.999969 & 0.0096540 \\
0.620000 & 0.89387 & 0.991290 & 0.16150 & 0.999972 & 0.0091140 \\
0.640000 & 0.87957 & 0.992238 & 0.15252 & 0.999975 & 0.0086042 \\
0.653263 & 0.86939 & 0.993082 & 0.14404 & 0.999978 & 0.0081229 \\
0.660000 & 0.86401 & 0.993834 & 0.13602 & 0.999981 & 0.0076685 \\
0.680000 & 0.84710 & 0.994505 & 0.12845 & 0.999983 & 0.0072395 \\
0.690970 & 0.83723 & 0.995102 & 0.12129 & 0.999985 & 0.0068346 \\
0.700000 & 0.82877 & 0.995635 & 0.11453 & 0.999986 & 0.0064523 \\
0.720000 & 0.80890 & 0.996110 & 0.10815 & 0.999988 & 0.0060913 \\
0.724577 & 0.80412 & 0.996533 & 0.10212 & 0.999989 & 0.0057506 \\
0.740000 & 0.78735 & 0.996910 & 0.096419 & 1 & 0 \\
0.754529 & 0.77057 \\
	\end{tabular}
	\end{ruledtabular}
\end{table*}

\begin{table*}
\caption{Normalized superconducting state electronic entropy $S_{\rm es}/\gamma_{\rm n}T_{\rm c}$ versus reduced temperature $t = T/T_{\rm c}$ predicted by the $\alpha$-model in Eq.~(\ref{Eq:Se6}) for the $\alpha$ values listed, where $\alpha = \Delta(0)/k_{\rm B}T$\@. The exact value of $\alpha_{\rm BCS}\approx 1.764$ is given in Eq.~(\ref{Eq:alphaBCS}).}
\label{Tab:AlphaModelSes}
\begin{ruledtabular}
\begin{tabular}{c c c c c c c c}
&&&&$S_{\rm es}/\gamma_{\rm n}T_{\rm c}$\\
$t$ & $\alpha$ = 1 & $\alpha$ = 1.25 & $\alpha$ = 1.5 & $\alpha$ = $\alpha_{\rm BCS}$ & $\alpha$ = 2 & $\alpha$ = 2.25 & $\alpha$ = 2.5 \\
\hline
0.00 & 0 & 0 & 0 & 0 & 0 & 0 & 0 \\
0.02 & 1.0784e-21 & 5.5751e-27 & 2.7177e-32 & 6.4290e-38 & 5.7750e-43 & 2.5627e-48 & 1.1167e-53 \\
0.04 & 5.6943e-11 & 1.5142e-13 & 3.8053e-16 & 6.5731e-19 & 2.1568e-21 & 4.9478e-24 & 1.1150e-26 \\
0.06 & 2.0045e-07 & 4.2516e-09 & 8.5414e-11 & 1.3255e-12 & 3.1042e-14 & 5.7079e-16 & 1.0314e-17 \\
0.08 & 1.1596e-05 & 6.9241e-07 & 3.9244e-08 & 1.8224e-09 & 1.1389e-10 & 5.9230e-12 & 3.0284e-13 \\
0.10 & 1.3076e-04 & 1.4495e-05 & 1.5281e-06 & 1.3680e-07 & 1.5391e-08 & 1.4926e-09 & 1.4236e-10 \\
0.12 & 6.5353e-04 & 1.0923e-04 & 1.7394e-05 & 2.4094e-06 & 4.0091e-07 & 5.8866e-08 & 8.5033e-09 \\
0.14 & 2.0559e-03 & 4.6012e-04 & 9.8269e-05 & 1.8577e-05 & 4.0856e-06 & 8.0636e-07 & 1.5662e-07 \\
0.16 & 4.8480e-03 & 1.3493e-03 & 3.5885e-04 & 8.5599e-05 & 2.3195e-05 & 5.7125e-06 & 1.3851e-06 \\
0.18 & 9.4393e-03 & 3.1107e-03 & 9.8060e-04 & 2.8013e-04 & 8.9251e-05 & 2.6103e-05 & 7.5185e-06 \\
0.20 & 1.6080e-02 & 6.0633e-03 & 2.1889e-03 & 7.2207e-04 & 2.6180e-04 & 8.7833e-05 & 2.9029e-05 \\
0.22 & 2.4863e-02 & 1.0465e-02 & 4.2201e-03 & 1.5656e-03 & 6.3083e-04 & 2.3676e-04 & 8.7560e-05 \\
0.24 & 3.5758e-02 & 1.6494e-02 & 7.2929e-03 & 2.9834e-03 & 1.3125e-03 & 5.4078e-04 & 2.1963e-04 \\
0.26 & 4.8656e-02 & 2.4247e-02 & 1.1590e-02 & 5.1499e-03 & 2.4403e-03 & 1.0881e-03 & 4.7835e-04 \\
0.28 & 6.3395e-02 & 3.3758e-02 & 1.7251e-02 & 8.2286e-03 & 4.1559e-03 & 1.9830e-03 & 9.3307e-04 \\
0.30 & 7.9798e-02 & 4.5005e-02 & 2.4372e-02 & 1.2364e-02 & 6.6000e-03 & 3.3400e-03 & 1.6672e-03 \\
0.32 & 9.7683e-02 & 5.7933e-02 & 3.3012e-02 & 1.7678e-02 & 9.9064e-03 & 5.2791e-03 & 2.7753e-03 \\
0.34 & 1.1688e-01 & 7.2463e-02 & 4.3197e-02 & 2.4267e-02 & 1.4198e-02 & 7.9208e-03 & 4.3600e-03 \\
0.36 & 1.3722e-01 & 8.8503e-02 & 5.4927e-02 & 3.2209e-02 & 1.9585e-02 & 1.1383e-02 & 6.5286e-03 \\
0.38 & 1.5857e-01 & 1.0596e-01 & 6.8184e-02 & 4.1557e-02 & 2.6163e-02 & 1.5778e-02 & 9.3914e-03 \\
0.40 & 1.8079e-01 & 1.2472e-01 & 8.2937e-02 & 5.2352e-02 & 3.4016e-02 & 2.1214e-02 & 1.3059e-02 \\
0.42 & 2.0379e-01 & 1.4471e-01 & 9.9144e-02 & 6.4618e-02 & 4.3214e-02 & 2.7790e-02 & 1.7642e-02 \\
0.44 & 2.2746e-01 & 1.6582e-01 & 1.1676e-01 & 7.8367e-02 & 5.3818e-02 & 3.5599e-02 & 2.3249e-02 \\
0.46 & 2.5171e-01 & 1.8798e-01 & 1.3573e-01 & 9.3603e-02 & 6.5876e-02 & 4.4728e-02 & 2.9986e-02 \\
0.48 & 2.7649e-01 & 2.1110e-01 & 1.5600e-01 & 1.1032e-01 & 7.9431e-02 & 5.5256e-02 & 3.7957e-02 \\
0.50 & 3.0173e-01 & 2.3510e-01 & 1.7752e-01 & 1.2852e-01 & 9.4517e-02 & 6.7257e-02 & 4.7266e-02 \\
0.52 & 3.2736e-01 & 2.5993e-01 & 2.0024e-01 & 1.4817e-01 & 1.1116e-01 & 8.0801e-02 & 5.8011e-02 \\
0.54 & 3.5336e-01 & 2.8552e-01 & 2.2411e-01 & 1.6926e-01 & 1.2938e-01 & 9.5951e-02 & 7.0291e-02 \\
0.56 & 3.7968e-01 & 3.1181e-01 & 2.4907e-01 & 1.9178e-01 & 1.4921e-01 & 1.1277e-01 & 8.4202e-02 \\
0.58 & 4.0628e-01 & 3.3876e-01 & 2.7508e-01 & 2.1569e-01 & 1.7065e-01 & 1.3131e-01 & 9.9840e-02 \\
0.60 & 4.3314e-01 & 3.6631e-01 & 3.0209e-01 & 2.4099e-01 & 1.9371e-01 & 1.5162e-01 & 1.1730e-01 \\
0.62 & 4.6023e-01 & 3.9443e-01 & 3.3006e-01 & 2.6763e-01 & 2.1840e-01 & 1.7377e-01 & 1.3667e-01 \\
0.64 & 4.8753e-01 & 4.2307e-01 & 3.5895e-01 & 2.9561e-01 & 2.4474e-01 & 1.9779e-01 & 1.5805e-01 \\
0.66 & 5.1501e-01 & 4.5221e-01 & 3.8871e-01 & 3.2488e-01 & 2.7271e-01 & 2.2373e-01 & 1.8153e-01 \\
0.68 & 5.4267e-01 & 4.8180e-01 & 4.1932e-01 & 3.5544e-01 & 3.0234e-01 & 2.5164e-01 & 2.0720e-01 \\
0.70 & 5.7048e-01 & 5.1183e-01 & 4.5073e-01 & 3.8726e-01 & 3.3361e-01 & 2.8156e-01 & 2.3516e-01 \\
0.72 & 5.9843e-01 & 5.4226e-01 & 4.8292e-01 & 4.2030e-01 & 3.6654e-01 & 3.1354e-01 & 2.6551e-01 \\
0.74 & 6.2652e-01 & 5.7306e-01 & 5.1585e-01 & 4.5456e-01 & 4.0111e-01 & 3.4761e-01 & 2.9834e-01 \\
0.76 & 6.5473e-01 & 6.0423e-01 & 5.4949e-01 & 4.8999e-01 & 4.3734e-01 & 3.8382e-01 & 3.3375e-01 \\
0.78 & 6.8304e-01 & 6.3573e-01 & 5.8381e-01 & 5.2659e-01 & 4.7521e-01 & 4.2221e-01 & 3.7185e-01 \\
0.80 & 7.1146e-01 & 6.6755e-01 & 6.1880e-01 & 5.6434e-01 & 5.1473e-01 & 4.6282e-01 & 4.1273e-01 \\
0.82 & 7.3998e-01 & 6.9967e-01 & 6.5442e-01 & 6.0320e-01 & 5.5589e-01 & 5.0568e-01 & 4.5650e-01 \\
0.84 & 7.6858e-01 & 7.3208e-01 & 6.9064e-01 & 6.4315e-01 & 5.9870e-01 & 5.5085e-01 & 5.0327e-01 \\
0.86 & 7.9727e-01 & 7.6475e-01 & 7.2746e-01 & 6.8419e-01 & 6.4315e-01 & 5.9836e-01 & 5.5316e-01 \\
0.88 & 8.2603e-01 & 7.9769e-01 & 7.6484e-01 & 7.2628e-01 & 6.8923e-01 & 6.4826e-01 & 6.0628e-01 \\
0.90 & 8.5487e-01 & 8.3086e-01 & 8.0277e-01 & 7.6941e-01 & 7.3695e-01 & 7.0057e-01 & 6.6276e-01 \\
0.92 & 8.8377e-01 & 8.6427e-01 & 8.4123e-01 & 8.1356e-01 & 7.8631e-01 & 7.5536e-01 & 7.2271e-01 \\
0.94 & 9.1274e-01 & 8.9789e-01 & 8.8020e-01 & 8.5871e-01 & 8.3729e-01 & 8.1265e-01 & 7.8627e-01 \\
0.96 & 9.4177e-01 & 9.3173e-01 & 9.1966e-01 & 9.0484e-01 & 8.8990e-01 & 8.7249e-01 & 8.5358e-01 \\
0.98 & 9.7086e-01 & 9.6577e-01 & 9.5960e-01 & 9.5195e-01 & 9.4414e-01 & 9.3492e-01 & 9.2477e-01 \\
1.00 & 1 & 1 & 1 & 1 & 1 & 1 & 1 \\
	\end{tabular}
	\end{ruledtabular}
\end{table*}

\begin{table*}
\caption{Normalized superconducting state electronic heat capacity $C_{\rm es}/\gamma_{\rm n}T_{\rm c}$ versus reduced temperature $t = T/T_{\rm c}$ predicted by the $\alpha$-model in Eq.~(\ref{Eq:Ces2}) for the $\alpha$ values listed, where $\alpha = \Delta(0)/k_{\rm B}T$\@. The exact value of $\alpha_{\rm BCS}\approx 1.764$ is given in Eq.~(\ref{Eq:alphaBCS}).}
\label{Tab:AlphaModelCes}
\begin{ruledtabular}
\begin{tabular}{c c c c c c c c}
&&&&$C_{\rm es}/\gamma_{\rm n}T_{\rm c}$\\
$t$ & $\alpha$ = 1 & $\alpha$ = 1.25 & $\alpha$ = 1.5 & $\alpha$ = $\alpha_{\rm BCS}$ & $\alpha$ = 2 & $\alpha$ = 2.25 & $\alpha$ = 2.5 \\
\hline
0.00 & 0 & 0 & 0 & 0 & 0 & 0 & 0 \\
0.02 & 5.3420e-20 & 3.4582e-25 & 2.0254e-30 & 5.6392e-36 & 5.7472e-41 & 2.8707e-46 & 1.3904e-51 \\
0.04 & 1.3992e-09 & 4.6649e-12 & 1.4098e-14 & 2.8684e-17 & 1.0684e-19 & 2.7600e-22 & 6.9165e-25 \\
0.06 & 3.2619e-06 & 8.6815e-08 & 2.0988e-09 & 3.8387e-11 & 1.0209e-12 & 2.1147e-14 & 4.2505e-16 \\
0.08 & 1.4076e-04 & 1.0551e-05 & 7.1992e-07 & 3.9418e-08 & 2.7984e-09 & 1.6400e-10 & 9.3298e-12 \\
0.10 & 1.2644e-03 & 1.7595e-04 & 2.2336e-05 & 2.3583e-06 & 3.0150e-07 & 3.2957e-08 & 3.4980e-09 \\
0.12 & 5.2492e-03 & 1.1011e-03 & 2.1114e-04 & 3.4498e-05 & 6.5239e-06 & 1.0799e-06 & 1.7363e-07 \\
0.14 & 1.4121e-02 & 3.9645e-03 & 1.0194e-03 & 2.2733e-04 & 5.6825e-05 & 1.2645e-05 & 2.7342e-06 \\
0.16 & 2.9088e-02 & 1.0151e-02 & 3.2496e-03 & 9.1427e-04 & 2.8158e-04 & 7.8195e-05 & 2.1108e-05 \\
0.18 & 5.0293e-02 & 2.0771e-02 & 7.8790e-03 & 2.6544e-03 & 9.6117e-04 & 3.1697e-04 & 1.0165e-04 \\
0.20 & 7.7077e-02 & 3.6407e-02 & 1.5811e-02 & 6.1494e-03 & 2.5338e-03 & 9.5848e-04 & 3.5270e-04 \\
0.22 & 1.0838e-01 & 5.7117e-02 & 2.7699e-02 & 1.2113e-02 & 5.5462e-03 & 2.3469e-03 & 9.6633e-04 \\
0.24 & 1.4304e-01 & 8.2570e-02 & 4.3895e-02 & 2.1163e-02 & 1.0578e-02 & 4.9137e-03 & 2.2218e-03 \\
0.26 & 1.8001e-01 & 1.1221e-01 & 6.4473e-02 & 3.3758e-02 & 1.8173e-02 & 9.1348e-03 & 4.4707e-03 \\
0.28 & 2.1843e-01 & 1.4540e-01 & 8.9299e-02 & 5.0188e-02 & 2.8795e-02 & 1.5487e-02 & 8.1128e-03 \\
0.30 & 2.5761e-01 & 1.8149e-01 & 1.1811e-01 & 7.0594e-02 & 4.2806e-02 & 2.4419e-02 & 1.3569e-02 \\
0.32 & 2.9707e-01 & 2.1990e-01 & 1.5056e-01 & 9.4995e-02 & 6.0474e-02 & 3.6326e-02 & 2.1261e-02 \\
0.34 & 3.3648e-01 & 2.6012e-01 & 1.8630e-01 & 1.2333e-01 & 8.1975e-02 & 5.1554e-02 & 3.1593e-02 \\
0.36 & 3.7561e-01 & 3.0174e-01 & 2.2496e-01 & 1.5547e-01 & 1.0742e-01 & 7.0389e-02 & 4.4951e-02 \\
0.38 & 4.1433e-01 & 3.4441e-01 & 2.6622e-01 & 1.9127e-01 & 1.3685e-01 & 9.3070e-02 & 6.1690e-02 \\
0.40 & 4.5256e-01 & 3.8785e-01 & 3.0975e-01 & 2.3054e-01 & 1.7030e-01 & 1.1980e-01 & 8.2141e-02 \\
0.42 & 4.9028e-01 & 4.3186e-01 & 3.5530e-01 & 2.7313e-01 & 2.0773e-01 & 1.5073e-01 & 1.0661e-01 \\
0.44 & 5.2747e-01 & 4.7626e-01 & 4.0262e-01 & 3.1884e-01 & 2.4911e-01 & 1.8601e-01 & 1.3539e-01 \\
0.46 & 5.6416e-01 & 5.2093e-01 & 4.5150e-01 & 3.6752e-01 & 2.9441e-01 & 2.2575e-01 & 1.6875e-01 \\
0.48 & 6.0037e-01 & 5.6576e-01 & 5.0176e-01 & 4.1899e-01 & 3.4356e-01 & 2.7006e-01 & 2.0696e-01 \\
0.50 & 6.3613e-01 & 6.1069e-01 & 5.5325e-01 & 4.7313e-01 & 3.9651e-01 & 3.1903e-01 & 2.5026e-01 \\
0.52 & 6.7147e-01 & 6.5566e-01 & 6.0582e-01 & 5.2978e-01 & 4.5320e-01 & 3.7275e-01 & 2.9891e-01 \\
0.54 & 7.0643e-01 & 7.0062e-01 & 6.5937e-01 & 5.8882e-01 & 5.1358e-01 & 4.3129e-01 & 3.5315e-01 \\
0.56 & 7.4104e-01 & 7.4556e-01 & 7.1380e-01 & 6.5014e-01 & 5.7759e-01 & 4.9473e-01 & 4.1324e-01 \\
0.58 & 7.7533e-01 & 7.9044e-01 & 7.6900e-01 & 7.1363e-01 & 6.4518e-01 & 5.6316e-01 & 4.7943e-01 \\
0.60 & 8.0934e-01 & 8.3525e-01 & 8.2492e-01 & 7.7919e-01 & 7.1630e-01 & 6.3665e-01 & 5.5198e-01 \\
0.62 & 8.4309e-01 & 8.7999e-01 & 8.8149e-01 & 8.4674e-01 & 7.9091e-01 & 7.1528e-01 & 6.3115e-01 \\
0.64 & 8.7660e-01 & 9.2465e-01 & 9.3865e-01 & 9.1618e-01 & 8.6897e-01 & 7.9914e-01 & 7.1724e-01 \\
0.66 & 9.0991e-01 & 9.6922e-01 & 9.9634e-01 & 9.8745e-01 & 9.5045e-01 & 8.8832e-01 & 8.1051e-01 \\
0.68 & 9.4303e-01 & 1.0137 & 1.0545 & 1.0605 & 1.0353 & 9.8289e-01 & 9.1129e-01 \\
0.70 & 9.7598e-01 & 1.0581 & 1.1132 & 1.1352 & 1.1235 & 1.0830 & 1.0199 \\
0.72 & 1.0088 & 1.1024 & 1.1722 & 1.2115 & 1.2150 & 1.1886 & 1.1366 \\
0.74 & 1.0414 & 1.1467 & 1.2317 & 1.2894 & 1.3098 & 1.3000 & 1.2618 \\
0.76 & 1.0740 & 1.1908 & 1.2915 & 1.3689 & 1.4078 & 1.4171 & 1.3959 \\
0.78 & 1.1064 & 1.2349 & 1.3517 & 1.4498 & 1.5092 & 1.5402 & 1.5393 \\
0.80 & 1.1387 & 1.2789 & 1.4122 & 1.5321 & 1.6137 & 1.6693 & 1.6923 \\
0.82 & 1.1710 & 1.3229 & 1.4730 & 1.6159 & 1.7214 & 1.8045 & 1.8554 \\
0.84 & 1.2031 & 1.3668 & 1.5341 & 1.7010 & 1.8324 & 1.9460 & 2.0291 \\
0.86 & 1.2352 & 1.4106 & 1.5954 & 1.7874 & 1.9465 & 2.0939 & 2.2139 \\
0.88 & 1.2673 & 1.4544 & 1.6570 & 1.8750 & 2.0638 & 2.2483 & 2.4102 \\
0.90 & 1.2992 & 1.4981 & 1.7189 & 1.9639 & 2.1842 & 2.4094 & 2.6186 \\
0.92 & 1.3311 & 1.5418 & 1.7809 & 2.0541 & 2.3078 & 2.5774 & 2.8398 \\
0.94 & 1.3630 & 1.5855 & 1.8432 & 2.1454 & 2.4345 & 2.7523 & 3.0743 \\
0.96 & 1.3948 & 1.6291 & 1.9057 & 2.2378 & 2.5644 & 2.9343 & 3.3229 \\
0.98 & 1.4266 & 1.6727 & 1.9684 & 2.3314 & 2.6974 & 3.1237 & 3.5861 \\
1.00 & 1.4584 & 1.7162 & 2.0313 & 2.4261 & 2.8335 & 3.3205 & 3.8649 \\
	\end{tabular}
	\end{ruledtabular}
\end{table*}

\begin{table*}
\caption{Thermodynamic critical field normalized to its zero temperature value,  $H_{\rm c}(t)/H_{\rm c}(0)$, versus reduced temperature $t = T/T_{\rm c}$ predicted by the $\alpha$-model in Eq.~(\ref{Eq:DeltaFe}) for the $\alpha$ values listed, where $\alpha = \Delta(0)/k_{\rm B}T$\@. The exact value of $\alpha_{\rm BCS}\approx 1.764$ is given in Eq.~(\ref{Eq:alphaBCS}) and the expression for $H_{\rm c}(0)$ versus $\alpha$ is given in Eq.~(\ref{Eq:AlphaModelHc0}).}
\label{Tab:AlphaModelHcH0}
\begin{ruledtabular}
\begin{tabular}{c c c c c c c c}
&&&&$H_{\rm c}(t)/H_{\rm c}(0)$\\
$t$ & $\alpha$ = 1 & $\alpha$ = 1.25 & $\alpha$ = 1.5 & $\alpha$ = $\alpha_{\rm BCS}$ & $\alpha$ = 2 & $\alpha$ = 2.25 & $\alpha$ = 2.5 \\
\hline
0.00 & 1 & 1 & 1 & 1 & 1 & 1 & 1 \\
0.02 & 0.99868 & 0.99916 & 0.99942 & 0.99958 & 0.99967 & 0.99974 & 0.99979 \\
0.04 & 0.99472 & 0.99663 & 0.99766 & 0.99831 & 0.99868 & 0.99896 & 0.99916 \\
0.06 & 0.98809 & 0.99239 & 0.99472 & 0.99619 & 0.99704 & 0.99766 & 0.99810 \\
0.08 & 0.97872 & 0.98643 & 0.99060 & 0.99321 & 0.99472 & 0.99583 & 0.99663 \\
0.10 & 0.96659 & 0.97872 & 0.98527 & 0.98937 & 0.99174 & 0.99348 & 0.99472 \\
0.12 & 0.95172 & 0.96924 & 0.97872 & 0.98466 & 0.98809 & 0.99060 & 0.99239 \\
0.14 & 0.93431 & 0.95798 & 0.97094 & 0.97906 & 0.98375 & 0.98718 & 0.98963 \\
0.16 & 0.91468 & 0.94503 & 0.96192 & 0.97256 & 0.97871 & 0.98321 & 0.98642 \\
0.18 & 0.89326 & 0.93050 & 0.95167 & 0.96514 & 0.97296 & 0.97868 & 0.98275 \\
0.20 & 0.87047 & 0.91456 & 0.94025 & 0.95680 & 0.96646 & 0.97355 & 0.97860 \\
0.22 & 0.84673 & 0.89739 & 0.92769 & 0.94752 & 0.95919 & 0.96777 & 0.97391 \\
0.24 & 0.82238 & 0.87917 & 0.91406 & 0.93730 & 0.95110 & 0.96132 & 0.96863 \\
0.26 & 0.79768 & 0.86008 & 0.89943 & 0.92613 & 0.94218 & 0.95413 & 0.96271 \\
0.28 & 0.77286 & 0.84025 & 0.88387 & 0.91402 & 0.93237 & 0.94615 & 0.95609 \\
0.30 & 0.74806 & 0.81983 & 0.86744 & 0.90098 & 0.92167 & 0.93733 & 0.94870 \\
0.32 & 0.72340 & 0.79891 & 0.85020 & 0.88702 & 0.91004 & 0.92764 & 0.94050 \\
0.34 & 0.69895 & 0.77758 & 0.83221 & 0.87216 & 0.89748 & 0.91702 & 0.93142 \\
0.36 & 0.67475 & 0.75591 & 0.81352 & 0.85640 & 0.88397 & 0.90547 & 0.92142 \\
0.38 & 0.65082 & 0.73396 & 0.79418 & 0.83979 & 0.86951 & 0.89294 & 0.91047 \\
0.40 & 0.62718 & 0.71177 & 0.77422 & 0.82232 & 0.85410 & 0.87941 & 0.89853 \\
0.42 & 0.60384 & 0.68937 & 0.75368 & 0.80403 & 0.83774 & 0.86488 & 0.88556 \\
0.44 & 0.58077 & 0.66680 & 0.73261 & 0.78494 & 0.82044 & 0.84933 & 0.87154 \\
0.46 & 0.55798 & 0.64408 & 0.71102 & 0.76506 & 0.80221 & 0.83276 & 0.85646 \\
0.48 & 0.53545 & 0.62121 & 0.68895 & 0.74443 & 0.78305 & 0.81515 & 0.84028 \\
0.50 & 0.51316 & 0.59823 & 0.66642 & 0.72306 & 0.76298 & 0.79650 & 0.82299 \\
0.52 & 0.49110 & 0.57514 & 0.64346 & 0.70098 & 0.74200 & 0.77682 & 0.80459 \\
0.54 & 0.46926 & 0.55195 & 0.62008 & 0.67820 & 0.72014 & 0.75610 & 0.78505 \\
0.56 & 0.44761 & 0.52867 & 0.59631 & 0.65474 & 0.69740 & 0.73434 & 0.76437 \\
0.58 & 0.42615 & 0.50531 & 0.57216 & 0.63063 & 0.67380 & 0.71156 & 0.74254 \\
0.60 & 0.40486 & 0.48186 & 0.54766 & 0.60588 & 0.64934 & 0.68774 & 0.71954 \\
0.62 & 0.38373 & 0.45834 & 0.52281 & 0.58051 & 0.62405 & 0.66290 & 0.69537 \\
0.64 & 0.36274 & 0.43475 & 0.49763 & 0.55454 & 0.59793 & 0.63703 & 0.67002 \\
0.66 & 0.34188 & 0.41108 & 0.47213 & 0.52798 & 0.57101 & 0.61014 & 0.64347 \\
0.68 & 0.32115 & 0.38736 & 0.44633 & 0.50085 & 0.54328 & 0.58224 & 0.61573 \\
0.70 & 0.30053 & 0.36356 & 0.42024 & 0.47317 & 0.51477 & 0.55333 & 0.58678 \\
0.72 & 0.28002 & 0.33971 & 0.39387 & 0.44494 & 0.48548 & 0.52340 & 0.55661 \\
0.74 & 0.25960 & 0.31579 & 0.36723 & 0.41619 & 0.45543 & 0.49248 & 0.52522 \\
0.76 & 0.23927 & 0.29182 & 0.34033 & 0.38693 & 0.42463 & 0.46055 & 0.49259 \\
0.78 & 0.21901 & 0.26779 & 0.31317 & 0.35717 & 0.39308 & 0.42762 & 0.45871 \\
0.80 & 0.19884 & 0.24371 & 0.28578 & 0.32692 & 0.36081 & 0.39369 & 0.42358 \\
0.82 & 0.17873 & 0.21957 & 0.25814 & 0.29619 & 0.32781 & 0.35877 & 0.38717 \\
0.84 & 0.15868 & 0.19537 & 0.23028 & 0.26500 & 0.29411 & 0.32286 & 0.34949 \\
0.86 & 0.13869 & 0.17113 & 0.20220 & 0.23336 & 0.25970 & 0.28596 & 0.31050 \\
0.88 & 0.11875 & 0.14683 & 0.17391 & 0.20127 & 0.22461 & 0.24807 & 0.27021 \\
0.90 & 0.09886 & 0.12248 & 0.14541 & 0.16875 & 0.18883 & 0.20919 & 0.22859 \\
0.92 & 0.07902 & 0.09808 & 0.11670 & 0.13581 & 0.15238 & 0.16932 & 0.18563 \\
0.94 & 0.05921 & 0.07363 & 0.08781 & 0.10246 & 0.11526 & 0.12847 & 0.14131 \\
0.96 & 0.03944 & 0.04914 & 0.05872 & 0.06870 & 0.07749 & 0.08664 & 0.09561 \\
0.98 & 0.01971 & 0.02459 & 0.02945 & 0.03454 & 0.03906 & 0.04381 & 0.04851 \\
1.00 & 0 & 0 & 0 & 0 & 0 & 0 & 0 \\
	\end{tabular}
	\end{ruledtabular}
\end{table*}

\begin{table*}
\caption{Normalized London parameter $I(t) = 1 - \frac{\Lambda(t)}{\Lambda(0)}$ from Eq.~(\ref{Eq:LambdaI(t)}) and normalized London penetration depth $\frac{\Delta\lambda_{\rm L}(t)}{\lambda(0)}\equiv \frac{\lambda_{\rm L}(t)}{\lambda_{\rm L}(0)}-1 = \frac{1}{\sqrt{1-I(t)}} - 1$ from Eq.~(\ref{Eq:lambdaratio2}) versus reduced temperature $t = T/T_{\rm c}$ according to the BCS theory.  The data are the same in the $\alpha$-model.}
\label{Tab:BCSlambda}
\begin{ruledtabular}
\begin{tabular}{c c c c c c}
$t$ & $I(t)$ & $\Delta\lambda_{\rm L}(t)/\lambda_{\rm L}(0)$ & $t$ & $I(t)$ & $\Delta\lambda_{\rm L}(t)/\lambda_{\rm L}(0)$ \\
\hline
0.06 & 2.3513e-12 & 1.1757e-12 & 0.54 & 2.1335e-01 & 1.2748e-01 \\
0.07 & 1.4543e-10 & 7.2714e-11 & 0.55 & 2.2597e-01 & 1.3664e-01 \\
0.08 & 3.1803e-09 & 1.5902e-09 & 0.56 & 2.3889e-01 & 1.4624e-01 \\
0.09 & 3.4812e-08 & 1.7406e-08 & 0.57 & 2.5209e-01 & 1.5631e-01 \\
0.10 & 2.3490e-07 & 1.1745e-07 & 0.58 & 2.6557e-01 & 1.6687e-01 \\
0.11 & 1.1155e-06 & 5.5774e-07 & 0.59 & 2.7932e-01 & 1.7795e-01 \\
0.12 & 4.0716e-06 & 2.0358e-06 & 0.60 & 2.9333e-01 & 1.8957e-01 \\
0.13 & 1.2142e-05 & 6.0713e-06 & 0.61 & 3.0759e-01 & 2.0176e-01 \\
0.14 & 3.0902e-05 & 1.5451e-05 & 0.62 & 3.2211e-01 & 2.1456e-01 \\
0.15 & 6.9287e-05 & 3.4645e-05 & 0.63 & 3.3686e-01 & 2.2800e-01 \\
0.16 & 1.4018e-04 & 7.0099e-05 & 0.64 & 3.5185e-01 & 2.4211e-01 \\
0.17 & 2.6064e-04 & 1.3035e-04 & 0.65 & 3.6706e-01 & 2.5695e-01 \\
0.18 & 4.5173e-04 & 2.2594e-04 & 0.66 & 3.8249e-01 & 2.7256e-01 \\
0.19 & 7.3799e-04 & 3.6920e-04 & 0.67 & 3.9814e-01 & 2.8899e-01 \\
0.20 & 1.1467e-03 & 5.7386e-04 & 0.68 & 4.1399e-01 & 3.0631e-01 \\
0.21 & 1.7071e-03 & 8.5464e-04 & 0.69 & 4.3004e-01 & 3.2458e-01 \\
0.22 & 2.4491e-03 & 1.2268e-03 & 0.70 & 4.4629e-01 & 3.4387e-01 \\
0.23 & 3.4029e-03 & 1.7058e-03 & 0.71 & 4.6272e-01 & 3.6427e-01 \\
0.24 & 4.5978e-03 & 2.3069e-03 & 0.72 & 4.7934e-01 & 3.8587e-01 \\
0.25 & 6.0616e-03 & 3.0447e-03 & 0.73 & 4.9613e-01 & 4.0878e-01 \\
0.26 & 7.8204e-03 & 3.9333e-03 & 0.74 & 5.1310e-01 & 4.3311e-01 \\
0.27 & 9.8978e-03 & 4.9859e-03 & 0.75 & 5.3023e-01 & 4.5901e-01 \\
0.28 & 1.2315e-02 & 6.2149e-03 & 0.76 & 5.4753e-01 & 4.8663e-01 \\
0.29 & 1.5090e-02 & 7.6316e-03 & 0.77 & 5.6498e-01 & 5.1616e-01 \\
0.30 & 1.8240e-02 & 9.2465e-03 & 0.78 & 5.8258e-01 & 5.4779e-01 \\
0.31 & 2.1776e-02 & 1.1069e-02 & 0.79 & 6.0033e-01 & 5.8179e-01 \\
0.32 & 2.5710e-02 & 1.3109e-02 & 0.80 & 6.1822e-01 & 6.1843e-01 \\
0.33 & 3.0051e-02 & 1.5373e-02 & 0.81 & 6.3625e-01 & 6.5805e-01 \\
0.34 & 3.4803e-02 & 1.7870e-02 & 0.82 & 6.5441e-01 & 7.0106e-01 \\
0.35 & 3.9972e-02 & 2.0606e-02 & 0.83 & 6.7270e-01 & 7.4795e-01 \\
0.36 & 4.5559e-02 & 2.3589e-02 & 0.84 & 6.9112e-01 & 7.9931e-01 \\
0.37 & 5.1565e-02 & 2.6824e-02 & 0.85 & 7.0966e-01 & 8.5587e-01 \\
0.38 & 5.7988e-02 & 3.0319e-02 & 0.86 & 7.2832e-01 & 9.1853e-01 \\
0.39 & 6.4827e-02 & 3.4080e-02 & 0.87 & 7.4709e-01 & 9.8846e-01 \\
0.40 & 7.2078e-02 & 3.8112e-02 & 0.88 & 7.6597e-01 & 1.0671 \\
0.41 & 7.9736e-02 & 4.2423e-02 & 0.89 & 7.8496e-01 & 1.1565 \\
0.42 & 8.7797e-02 & 4.7018e-02 & 0.90 & 8.0405e-01 & 1.2591 \\
0.43 & 9.6255e-02 & 5.1906e-02 & 0.91 & 8.2324e-01 & 1.3786 \\
0.44 & 1.0510e-01 & 5.7093e-02 & 0.92 & 8.4253e-01 & 1.5200 \\
0.45 & 1.1433e-01 & 6.2587e-02 & 0.93 & 8.6192e-01 & 1.6911 \\
0.46 & 1.2394e-01 & 6.8397e-02 & 0.94 & 8.8139e-01 & 1.9037 \\
0.47 & 1.3391e-01 & 7.4532e-02 & 0.95 & 9.0096e-01 & 2.1775 \\
0.48 & 1.4425e-01 & 8.1000e-02 & 0.96 & 9.2061e-01 & 2.5490 \\
0.49 & 1.5493e-01 & 8.7813e-02 & 0.97 & 9.4034e-01 & 3.0940 \\
0.50 & 1.6596e-01 & 9.4982e-02 & 0.98 & 9.6015e-01 & 4.0093 \\
0.51 & 1.7733e-01 & 1.0252e-01 & 0.99 & 9.8004e-01 & 6.0775 \\
0.52 & 1.8902e-01 & 1.1044e-01 & 1.00 & 1 \\
0.53 & 2.0103e-01 & 1.1875e-01 \\
	\end{tabular}
	\end{ruledtabular}
\end{table*}

\begin{table}
\caption{Normalized Pippard penetration depth $\frac{\Delta\lambda_{\rm P}(t)}{\lambda_{\rm P}(0)}$ from Eq.~(\ref{Eq:lambdaPippard}) versus reduced temperature $t = T/T_{\rm c}$ according to the BCS theory.}
\label{Tab:BCSlambdaPippard}
\begin{ruledtabular}
\begin{tabular}{c c c c}
$t$ & $\Delta\lambda_{\rm P}(t)/\lambda_{\rm P}(0)$ & $t$ & $\Delta\lambda_{\rm P}(t)/\lambda_{\rm P}(0)$ \\
\hline
0.16 & 1.4595e-05 & 0.58 & 7.1394e-02 \\
0.17 & 2.8455e-05 & 0.59 & 7.6631e-02 \\
0.18 & 5.1314e-05 & 0.60 & 8.2150e-02 \\
0.19 & 8.6852e-05 & 0.61 & 8.7964e-02 \\
0.20 & 1.3942e-04 & 0.62 & 9.4090e-02 \\
0.21 & 2.1399e-04 & 0.63 & 1.0054e-01 \\
0.22 & 3.1601e-04 & 0.64 & 1.0734e-01 \\
0.23 & 4.5135e-04 & 0.65 & 1.1451e-01 \\
0.24 & 6.2615e-04 & 0.66 & 1.2207e-01 \\
0.25 & 8.4673e-04 & 0.67 & 1.3005e-01 \\
0.26 & 1.1195e-03 & 0.68 & 1.3847e-01 \\
0.27 & 1.4509e-03 & 0.69 & 1.4736e-01 \\
0.28 & 1.8472e-03 & 0.70 & 1.5676e-01 \\
0.29 & 2.3148e-03 & 0.71 & 1.6671e-01 \\
0.30 & 2.8596e-03 & 0.72 & 1.7724e-01 \\
0.31 & 3.4876e-03 & 0.73 & 1.8842e-01 \\
0.32 & 4.2046e-03 & 0.74 & 2.0028e-01 \\
0.33 & 5.0161e-03 & 0.75 & 2.1290e-01 \\
0.34 & 5.9275e-03 & 0.76 & 2.2635e-01 \\
0.35 & 6.9441e-03 & 0.77 & 2.4069e-01 \\
0.36 & 8.0709e-03 & 0.78 & 2.5603e-01 \\
0.37 & 9.3129e-03 & 0.79 & 2.7247e-01 \\
0.38 & 1.0675e-02 & 0.80 & 2.9014e-01 \\
0.39 & 1.2162e-02 & 0.81 & 3.0918e-01 \\
0.40 & 1.3780e-02 & 0.82 & 3.2977e-01 \\
0.41 & 1.5531e-02 & 0.83 & 3.5211e-01 \\
0.42 & 1.7423e-02 & 0.84 & 3.7645e-01 \\
0.43 & 1.9459e-02 & 0.85 & 4.0309e-01 \\
0.44 & 2.1645e-02 & 0.86 & 4.3240e-01 \\
0.45 & 2.3985e-02 & 0.87 & 4.6486e-01 \\
0.46 & 2.6486e-02 & 0.88 & 5.0106e-01 \\
0.47 & 2.9153e-02 & 0.89 & 5.4177e-01 \\
0.48 & 3.1991e-02 & 0.90 & 5.8800e-01 \\
0.49 & 3.5007e-02 & 0.91 & 6.4115e-01 \\
0.50 & 3.8208e-02 & 0.92 & 7.0315e-01 \\
0.51 & 4.1601e-02 & 0.93 & 7.7683e-01 \\
0.52 & 4.5192e-02 & 0.94 & 8.6653e-01 \\
0.53 & 4.8989e-02 & 0.95 & 9.7930e-01 \\
0.54 & 5.3002e-02 & 0.96 & 1.1277 \\
0.55 & 5.7238e-02 & 0.97 & 1.3370 \\
0.56 & 6.1709e-02 & 0.98 & 1.6697 \\
0.57 & 6.6423e-02 & 0.99 & 2.3567 \\
	\end{tabular}
	\end{ruledtabular}
\end{table}

%\clearpage

\end{document}